\documentclass[a4paper,11pt]{article}
\pdfoutput=1
\usepackage{jinstpub} 

\usepackage{xspace}
\usepackage[output-decimal-marker={.}, separate-uncertainty=true]{siunitx}
\usepackage{multirow}
\usepackage{booktabs}
\usepackage{xcolor}
\usepackage{subcaption}
\usepackage[noabbrev]{cleveref}
\usepackage[section]{placeins}
\usepackage{physics}
\usepackage{lineno}

\graphicspath{{images/}}
\newcommand{\uB}{MicroBooNE\xspace}
\newcommand{\cor}{CORSIKA\xspace}
\newcommand{\lartf}{LArTF\xspace}
\newcommand{\systs}{systematic uncertainties\xspace}
\newcommand{\syst}{systematic uncertainty\xspace}
\newcommand{\pct}[1]{$#1\%$}

\DeclareSIUnit\Hzz{events\per\s}
\title{Measurement of the Atmospheric Muon Rate with the MicroBooNE Liquid Argon TPC}

\author[jj]{P.~Abratenko}
\author[o]{M.~Alrashed}
\author[n]{R.~An}
\author[d]{J.~Anthony}
\author[ii]{J.~Asaadi}
\author[s]{A.~Ashkenazi}
\author[mm]{S.~Balasubramanian}
\author[k]{B.~Baller}
\author[t]{C.~Barnes}
\author[x]{G.~Barr}
% ...

% ...
\affiliation[d]{University of Cambridge, Cambridge CB3 0HE, United Kingdom}
% ...
\affiliation[k]{Fermi National Accelerator Laboratory (FNAL), Batavia, IL 60510, USA}
% ...
\affiliation[n]{Illinois Institute of Technology (IIT), Chicago, IL 60616, USA}
\affiliation[o]{Kansas State University (KSU), Manhattan, KS, 66506, USA}
% ...
\affiliation[s]{Massachusetts Institute of Technology (MIT), Cambridge, MA, 02139, USA}
\affiliation[t]{University of Michigan, Ann Arbor, MI, 48109, USA}
% ...
\affiliation[x]{University of Oxford, Oxford OX1 3RH, United Kingdom}
% ...
\affiliation[ii]{University of Texas, Arlington, TX, 76019, USA}
\affiliation[jj]{Tufts University, Medford, MA, 02155, USA}
% ...
\affiliation[mm]{Wright Laboratory, Department of Physics, Yale University, New Haven, CT, 06520, USA}

  \emailAdd{microboone\_info@fnal.gov}
\date{}
\abstract{
\uB is a near-surface liquid argon (LAr) time projection chamber (TPC) located at Fermilab. We measure muons originating from cosmic interactions in the atmosphere using both the charge collection and light readout detectors. The data is compared with the \cor cosmic-ray simulation. Agreement is found between the observation, simulation and previous results. Furthermore, the angular resolution of the reconstructed muons inside the TPC is studied in simulation.}
\arxivnumber{1234.56789} % only if you have one

\begin{document}
\maketitle
\flushbottom

\section{Introduction}
\label{ch:intro}

The \uB experiment is a neutrino detector located at Fermilab. Its main detector component is a Liquid Argon Time Projection Chamber (LArTPC) placed inside a cylindrical cryostat. Charged particles ionise the liquid argon while traversing the TPC. These ionisation electrons drift under the influence of an electric field and are sensed by three wire planes. The collection plane wires are vertical, while the wires on the two induction planes are offset by $+\ang{60}$ and $-\ang{60}$ with respect to the vertical. \uB is further equipped with an optical system of 32 units that provides an event trigger. Each optical unit consists of a 8-inch diameter Hamamatsu R5912-02mod cryogenic photomultiplier tube (PMT) inside a mu-metal shield. The PMTs are positioned behind individual tetraphenyl butadiene (TPB) coated acrylic plates to shift the VUV scintillation light created by interactions in the cryostat. A detailed description of the MicroBooNE detector can be found in~\cite{det}.

The \uB experiment is subject to a large atmospheric-muon flux due to its near-surface location. %~\cite{ub_cosmics_note}. 
The centre of the TPC is located \SI{6}{\m} below the surface in a concrete open pit in the liquid argon test facility building (LArTF). The large TPC dimensions -- the width ($x$) of \SI{2.6}{\m}, the height ($y$) of \SI{2.3}{\m} and the length ($z$) of \SI{10.4}{\m} -- lead to a large rate of cosmic particles. 

The TPC operating voltage of \SI{-70}{\kilo \V} is applied to the cathode, producing an electric field between the cathode and the anode in the $x$-direction. Therefore, the TPC charge collection time is approximately $\SI{2.3}{\ms}$, long enough to accumulate several atmospheric muons passing through the TPC. The majority of recorded events will consist of cosmic-ray activity, resulting in an important background for neutrino interaction measurements. 

In MicroBooNE, the main method we use to estimate the impact of cosmic activity on neutrino physics relies on data recorded when the neutrino beam is turned off. This is done in two configurations. First, the external data stream records events that are triggered by optical signals from cosmic activity in an identical way as the neutrino data stream. Data taken with the external data stream is used to account for events triggered by cosmic activity when the beam was on but no neutrino interacted. Second, an unbiased data stream records cosmic data in the usual event format without the optical trigger requirements. This data can then be overlaid with simulated neutrino activity. When correctly scaled to the number of protons on target, these overlay events can be combined with events from the external data stream and compared with beam-on data. These beam-off data streams can only be collected when the detector is operational. Even after several years of data-taking, the amount of beam-off data in both of these configurations is limited by operational constraints. Therefore, either for background studies before the experiment was fully commissioned, or for increased background statistics, accurate simulation of cosmic activity is essential. 

This work uses the unbiased data stream to validate the accuracy of simulated cosmic events. The simulation is scaled to the data using the exposure time. The results presented guide the simulation of cosmic activity for future near-surface experiments at Fermilab such as Mu2e, SBND, ICARUS and the DUNE prototype detectors at CERN~\cite{mu2e,sbn,dunend}. The cosmic-ray Monte Carlo (MC) simulation and its configuration are presented in \cref{ch:sim}.

LArTPCs are novel and complex detectors. In \cref{ch:tpc}, the \uB TPC is used to study the tracks created by atmospheric muons. Detector effects are discussed and we demonstrate that these are well modelled and understood. The optical system of MicroBooNE can be used to count muons, as is described in \cref{ch:pmt}. 

Independently of the importance for neutrino searches, the first atmospheric muon rate measurement with a surface-based liquid argon detector at Fermilab is presented. In \cref{ch:res}, the observed rates in the TPC and PMT systems are converted into an integrated atmospheric muon rate at the MicroBooNE TPC and above the roof of the building.  The measurements from the two independent systems are found to agree with one another. 

\section{Simulation of Cosmic Activity in \uB}
\label{ch:sim}

Cosmic rays are produced when galactic protons -- or heavier elements such as helium and up to iron -- interact with the Earth's atmosphere. These interactions lead to extensive air showers that contain a large number of cosmic-induced particles, here referred to as primaries. The composition and the flux of these particles depends on the location -- latitude, longitude and altitude -- as well as the amount of shielding above the detector.

\subsection{Geometry} 

\begin{figure}[htb]
	\centering
	\includegraphics[width=0.6\textwidth]{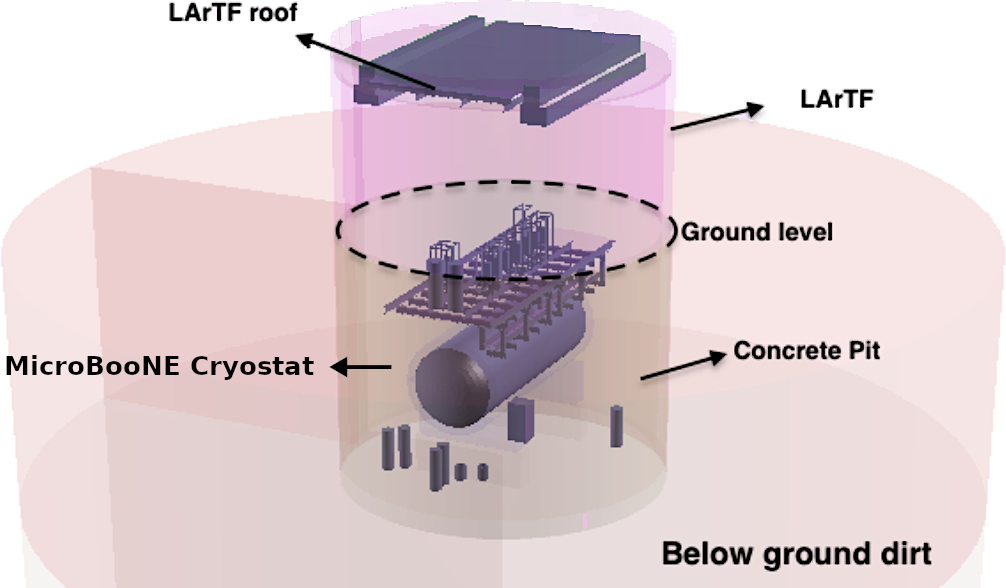}
	\caption{Geometry of the simulated MicroBooNE detector and surroundings. In order to show the details of the simulation, a transparent view of the \lartf geometry is shown.%Figure from~\cite{ub_cosmics_note}.
	}
	\label{fig:lartf}
\end{figure}

Fermilab is located at a latitude of \ang{41;50;15} North and a longitude of \ang{88;16;10} West. The \uB TPC is located inside an open concrete pit, \SI{6}{\m} underground and is at an elevation of \SI{228}{\m} above mean sea level.  The \uB simulated geometry includes the full \lartf building, with the roof and the concrete pit as well as the detector cryostat. The platform and electronics racks located on top of the cryostat are also included as shown in \cref{fig:lartf}. Muons are mainly affected by the concrete elements (density of \SI{2.3}{\g\per\cm^3}) and the dirt surrounding the  building (density of \SI{1.7}{\g\per\cm^3}).

\subsection{Cosmic-Ray Generators}
\label{sec:model}

The cosmic-ray simulation used by MicroBooNE is \cor version 7.4003~\cite{corsika}. Comparisons between this generator and  CRY -- another Monte-Carlo-based cosmic ray simulation library~\cite{cry} -- were performed. % in~\cite{ub_cosmics_note}.
Both generators predict similar muon rates. \cor was chosen because of the additional flexibility offered, such as the inclusion of longitudinal effects due to the movement of the Earth in the Milky Way~\cite{comptongetting} and a modifiable incoming flux of extra-terrestrial particles. 

The production of cosmic-ray particles in the Earth's atmosphere depends on the intensity of nucleons per energy-per-nucleon. The energy of the nucleons is approximately independent of whether the incident nucleons are free protons or are bound in heavier nuclei. The intensity of cosmic nucleons can be modelled by~\cite{pdg}: 
\begin{equation}
\Phi(E) = 1.8\times 10^4 (E[\si{\GeV}])^{-2.7} \: \: \frac{\si{nucleons}}{\si{\m^2 \: \s \: sr \: \GeV}},
\label{eq:flux}
\end{equation}
where $E$ is the energy-per-nucleon. In this work, in the default configuration, all galactic particles are modelled as cosmic protons. %In reality, approximately \pct{74} of extra-terrestrial nucleons are protons, \pct{18} are bound in helium nuclei~\cite{pamela} and the remainder largely consists of carbon and oxygen nuclei contributions~\cite{primarydatabase}. 
Especially at higher energies, components such as helium nuclei~\cite{pamela} and, in smaller fractions carbon, oxygen and even heavier nuclei contribute~\cite{primarydatabase}. 

An alternative \cor model of the extra-terrestrial flux, the constant mass composition (CMC) model, is used for comparison. While maintaining the same spectral index of $2.7$, this model has strong contributions from five mass groups: $A=1$ for protons, $A=4$ for helium nuclei, $A=14$ for the CNO group, $A=28$ for the Mg-Si group, and $A=56$ for the Fe group, where $A$ is the average mass number~\cite{1990}.

Hadronic interactions in the air showers are modelled by FLUKA~\cite{fluka}. The default configuration is summarised in \cref{tab:conf}

\begin{table}[htb]
\centering
\caption{The configuration parameters used in the comparisons with measurements later in this article. This configuration will be referred to as \textit{CORSIKA default}. The magnetic components are obtained from the latitude and longitude using NOAA~\cite{noaa}.}
\label{tab:conf}
\begin{tabular}{@{}lc@{}}
\toprule
Parameter                    & Value            \\ \midrule
CORSIKA                      & v7.4003          \\
Hadronic interaction model   & FLUKA 2011       \\
Flux constant                & \num{1.8e4}      \\
Flux energy slope            & \num{-2.7}       \\
Elevation                    & \SI{228}{\m}     \\
Magnetic field north-component & \SI{19.066}{\micro\tesla} \\
Magnetic field vertical-component & \SI{50.628}{\micro\tesla} \\ \bottomrule
\end{tabular}
\end{table}

\subsection{Simulated Muon Rate}
\label{ss:sim_rate}

%Due to the steeply falling energy dependence of \cref{eq:flux} and the low amount of shielding of the \uB experiment, nucleon energies in the range of \SIrange{0.5}{10}{\GeV}
 The simulated muon rate, $I [\si{\Hzz \per \m^2}]$, is obtained after integrating the muon flux over the muon energy and solid angle. This quantity is related to the often quoted integral intensity of vertical muons $I_v [\si{\Hzz \per \m^2 \per sr}]$ using the zenith angle distribution, $f(\theta)$:
\begin{equation}
    I [\si{\Hzz \per \m^2}] = \int_\Omega \int_E f(\theta) \; I_{v}(E) \; \text{d}\Omega \text{d}E, 
    \label{eq:int_intensity}
\end{equation}
where it is assumed that the muon rate can be factorised in the muon energy, the azimuthal angle and the zenith angle. In the case of azimuthal symmetry this reduces to:
\begin{equation}
    I [\si{\Hzz \per \m^2}] = 2\pi I_{v} \int_0^{\pi/2} f(\theta) \sin\theta \; \text{d}\theta.
\end{equation}
Often used analytical approximations for $f(\theta)$ are $\cos^2\theta$ and $\exp(1-1/\cos\theta)$~\cite{grieder}.

The propagation of the primary particles produced by the cosmic-ray generator inside the MicroBooNE environment is performed by the Geant4 program~\cite{g4}. The muon trajectory is described by a set of interaction points with corresponding energy losses. Due to the shielding provided by the walls of, and earth surrounding, \lartf and the liquid argon inside the cryostat surrounding the TPC, muons at the surface need a certain momentum to penetrate into the TPC. This momentum threshold depends on the angle of the incoming muon, ranging from $\gtrsim\SI{0.3}{\GeV \per c}$ for vertical muons to $\gtrsim\SI{1.5}{\GeV \per c}$ for muons with an incident angle of \ang{75} with the zenith. 

The trajectory of the simulated muon ends when the muon leaves the environment or stops inside the TPC. The physics processes simulated are different for $\mu^+$ and $\mu^-$. Muon decay is the only mechanism simulated for a stopped $\mu^+$ in the TPC, leading to a Michel positron. For stopped $\mu^-$, the dominant physics process %(\pct{63}) 
is nuclear absorption ($\mu^- + p \rightarrow n + \nu_\mu$), leaving no visible signature in the TPC. The remaining fraction of stopped $\mu^-$ in the TPC volume decay to Michel electrons. Overall, in simulation \pct{11.6\pm0.1} of the atmospheric muons entering the TPC stop in the TPC.

The atmospheric muon rate at the Earth's surface as predicted by different cosmic-ray simulations is given in \cref{tab:muons}. The rate of muons entering the MicroBooNE TPC volume, predicted with the default \cor configuration and obtained with the full detector simulation, is \SI{4388+-9}{\Hzz} (stat). Equivalently, this corresponds to $\order{10}$ atmospheric muons in a drift window of \SI{2.3}{\ms}, stressing the importance of understanding and modelling this background in near-surface LArTPCs.

\begin{table}[htb]
\caption[Simulated integrated atmospheric muon fluxes at \uB using different generators]
{Surface muon rates at the geomagnetic location of Fermilab with an altitude of \SI{228}{\m} for different cosmic-ray simulations. The rates are integrated over energies and angles. The quoted errors are statistical. The predicted rate with the CMC model is significantly higher than the default model, demonstrating the effect of the incoming flux parametrisation. The CMC model was chosen to provide an upper limit of the expected cosmic backgrounds in the MicroBooNE experiment.}
\label{tab:muons}
\centering
\small
\begin{tabular}{cccc}
\toprule
               & CRY &  \cor default &  \cor CMC \\ \midrule
$\mu^\pm$ flux [\si{\Hzz \per \m^2 }] \hfill & $120.0\pm0.2$ & $127.7\pm0.2$              & $160.9\pm0.3$                      \\ \bottomrule
\end{tabular}
\end{table}

\section{Atmospheric Muon Characterisation with the Time Projection Chamber}
\label{ch:tpc}

When a muon with an energy in the range \SIrange{0.1}{100}{\GeV} propagates through the liquid argon, it behaves approximately as a minimum ionising particle, depositing $\SI{\approx 2.1}{\MeV / \cm}$~\cite{pdg,ub_calibration} and follows a roughly straight trajectory. Small deflections at low muon energies $\lesssim \SI{1}{\GeV}$ arise from multiple Coulomb scattering~\cite{mcs}. Knock-on electrons, called $\delta$-rays, can have a high enough energy to appear as small side-tracks of the muon. In liquid argon, approximately two $\delta$-rays above \SI{4}{\MeV} are created every meter~\cite{deltaicarus,muonargon}. Muons decaying into a Michel electron or positron inside the TPC create a small electromagnetic shower with a visible energy peaked at \SI{\approx 20}{\MeV}~\cite{michel}. In the TPC, the position of atmospheric muons along the drift direction is unknown, therefore, it is not possible to distinguish a muon leaving the front or back face of the TPC from a stopping muon, unless the Michel decay is observed. Furthermore, because scattering, ionisation, and $\delta$-ray production do not depend strongly on the muon momentum, this momentum cannot be determined for muons above \SI{\approx 5}{\GeV}.

\begin{figure}
    \centering
    \begin{subfigure}{\textwidth}
    \centering
    \includegraphics[width=0.9\textwidth]{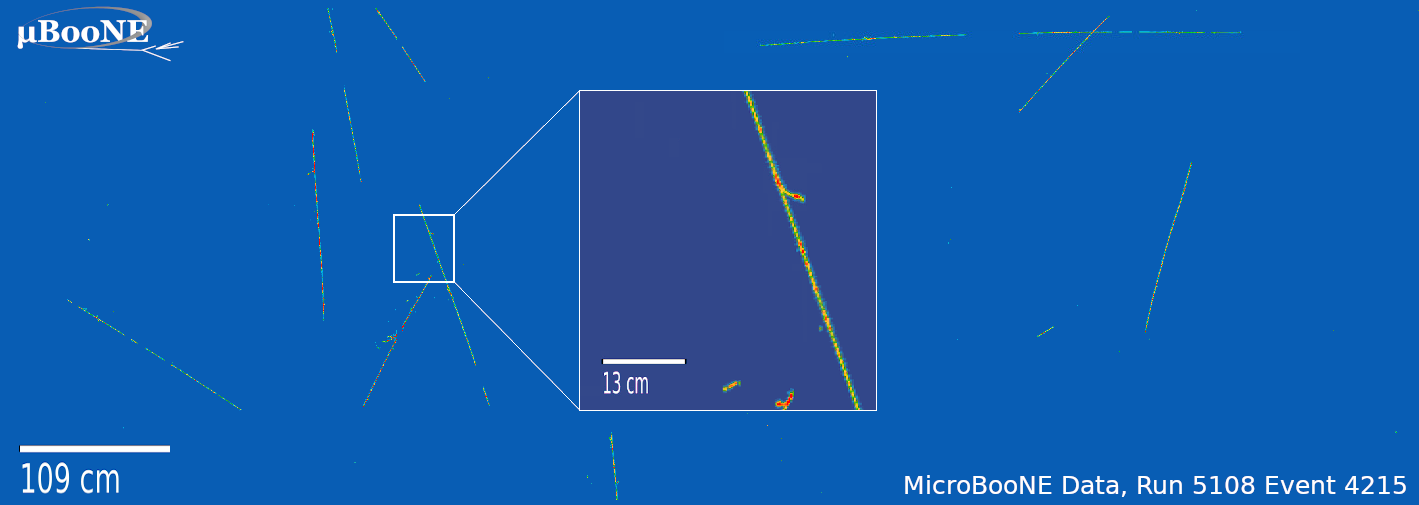}
    \caption{Wire waveform signals. The colour scale represents the amount of deposited charge.}
    \end{subfigure}
    \begin{subfigure}{\textwidth}
    \vspace{3mm}
    \centering
    \includegraphics[width=0.9\textwidth]{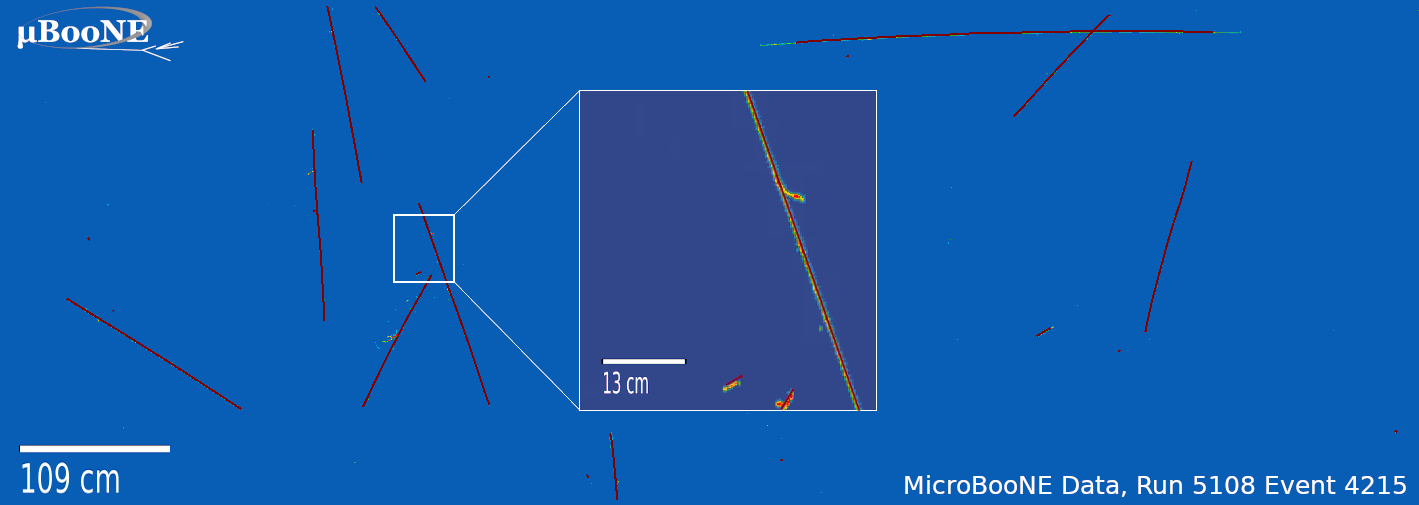}
    \caption{The same event is now overlaid with the reconstructed track objects in dark red.}
    \end{subfigure}
    \caption{Example event containing cosmic data as recorded on the collection plane. The horizontal axis corresponds to the wire number and the vertical axis to the charge collection time. 
    The total time is \SI{3.2}{\ms}, which includes the \SI{2.3}{\ms} that corresponds to one full drift width of the TPC. The white box zooms in on a fraction of the track to highlight the level of detail in the TPC, where a $\delta$-ray  is clearly visible.}
    \label{fig:ext_evd}
\end{figure}

\subsection{Space Charge Effect in LArTPC's}
\label{ss:sce}

To correctly reconstruct muon trajectories in the TPC, a good knowledge of the magnitude and direction of the electric field is crucial. The electric field is assumed to be uniform between the cathode and the anode plane. However, the continuous cosmic activity in combination with the slow drift time of the positive ions leads to a build-up of positive charge in the TPC, distorting the electric field. The effect of this distortion on the reconstructed position of collected ionisation electrons is referred to as the space charge effect~\cite{yifan}. The magnitude and the direction of this effect depends on the position of the deposited charge. Effectively, the majority of tracks inside the TPC have a reconstructed length that appears shorter than their simulated length.
The magnitude of the space charge effect is bigger for charge deposited far away from the collection planes (drift direction) and charge deposited near the edges in the plane perpendicular on the drift direction. The correction ranges up to $\order{\SI{10}{\cm}}$.

Throughout this work, two models are used and compared to take into account the space charge effect; a simulation-based model and a data-driven one. Both models are static approximations of the effect and ignore small time variations due to operational conditions.

The simulation-based model makes use of a Fourier series solution with boundary conditions to solve for the static electric field in a three-dimensional grid inside the TPC. The correction in the reconstructed position of ionisation electrons is obtained from the electric field solution using the Runge-Kutta-Fehlberg method (RKF45) for ray-tracing~\cite{mooney}.

The data-driven model uses an integrated UV laser system in combination with samples of through-going atmospheric muons. Both the laser and atmospheric muon methods enable MicroBooNE to further calibrate the residual space charge effects after applying the simulation-based model~\cite{mooney2}. 

In \cref{sc:tpc_datamc} both models will be compared with data and with the uncorrected simulation. The data-driven space charge effect correction is used as the default, the difference between the models is used to estimate a \syst in \cref{sc:tpc_syst}. Differences between data and simulation in the field response which could lead to biases in the reconstruction efficiency for either the track length or angles are discussed in \cite{signal1,signal2} and covered by the different space charge models.

\subsection{Muon Selection: Purity and Efficiency}
\label{ss:tpcpureff}

The straight muon trajectories are reconstructed as track objects using the cosmic reconstruction configuration of the Pandora reconstruction framework~\cite{pandora}. An example event display, as seen by the collection plane, before and after track reconstruction is shown in \cref{fig:ext_evd}. Tracks with a reconstructed track length longer than \SI{25}{\cm} are selected as muons and are used to study the purity and efficiency. Of simulated muons that intersect the TPC, \pct{6.8} have path lengths shorter than \SI{25}{\cm}. The selection purity is defined as the fraction of selected tracks for which the reconstructed charge is dominantly deposited by the simulated muon. Additionally, the reconstructed track needs to be uniquely associated to the simulated muon. The obtained purity for the selection is \pct{97.71\pm0.02} (stat).

Of the selected tracks, \pct{1.32\pm0.01} (stat) are not uniquely matched to a simulated muon. This is the case when a simulated muon is reconstructed as two separate tracks, for example, due to a scatter or the presence of an unresponsive wire region in the detector~\cite{pandora}. The remaining \pct{0.97\pm0.01} (stat) of the selected tracks are created by cosmogenic neutrons through inelastic scattering (in two thirds of the cases) and cosmogenic protons (in one third). 

The muon reconstruction efficiency, \pct{97.84\pm0.05} (stat), is defined as the fraction of simulated muons with a length of at least \SI{25}{\cm} inside the TPC that have a corresponding reconstructed track of at least \SI{25}{\cm}. The inefficiency of \pct{\approx 2} has two causes. First, \pct{\approx 10} of the charge read-out wires of the TPC are unresponsive~\cite{ub_noise}. A track with a significant portion of its trajectory lost due to unresponsive wires will not be reconstructed. Second, the reconstructed track length can be shorter than \SI{25}{\cm}, either because of unresponsive wires or of space-charge effects.

The reconstruction efficiency of \pct{97.84\pm0.05} (stat) is in agreement with a previous data-driven efficiency measurement using an external muon counter system that found $97.1\pm0.1~(\text{stat}) \pm 1.4~(\text{sys})\%$~\cite{mucs}. 

\subsection{Muon Reconstruction Performance in Simulation}
\label{sc:tpc_reso}

\begin{figure}[htb]
	\centering
	\includegraphics[width=0.8\textwidth]{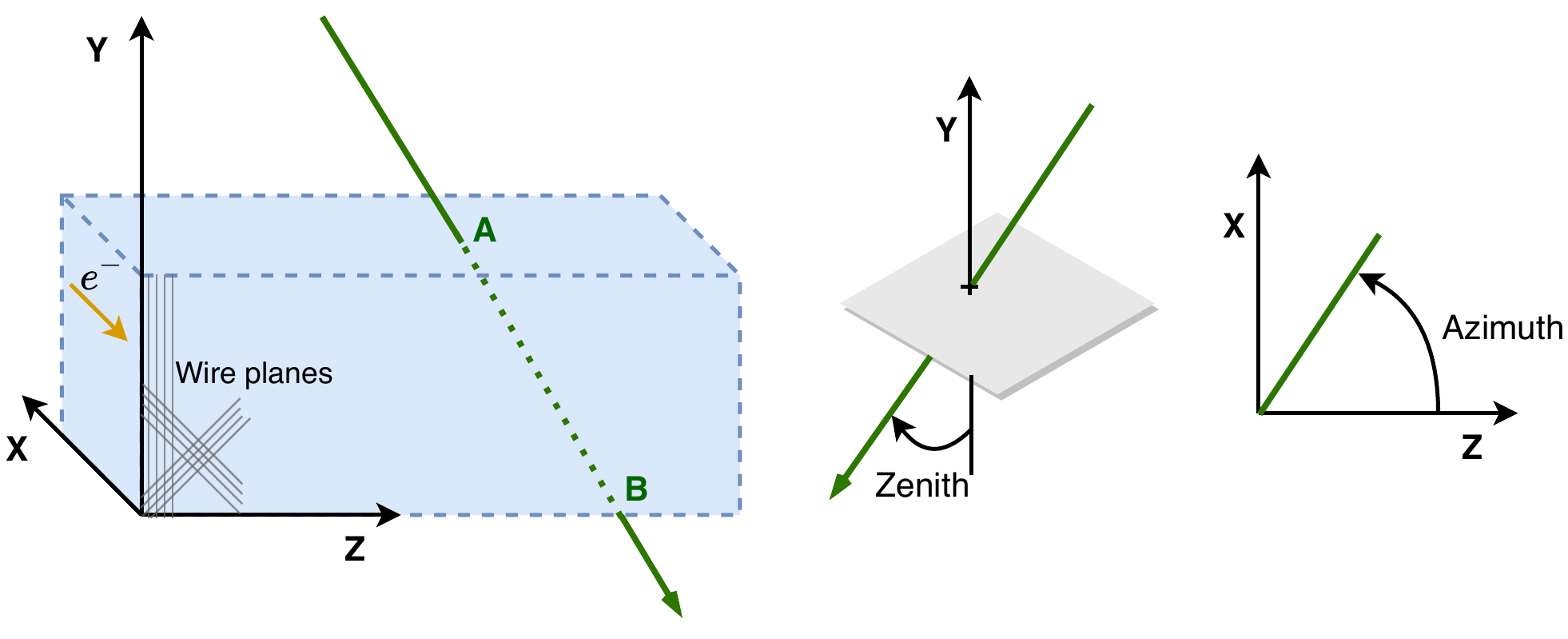}
	\caption[Coordinate frame and angle definitions for muon studies in the TPC]
	{Schematic representation of a muon (green) crossing the \uB TPC. The muon enters the TPC at point $A$ and exits the TPC, or stops inside it, at point $B$. The track length, azimuth and zenith angles are obtained from the segment $AB$. The left diagram illustrates the position of a few wires of each of the three planes (grey) and the direction of the drifting electrons (orange).}
	\label{fig:drawing}
\end{figure}

The reconstructed muon-track length and angles are defined in \cref{fig:drawing}. Before the in-TPC segment of the simulated muon trajectory is compared with the reconstructed track, a data-driven space charge effect correction is applied on the trajectory~\cite{mooney2}. 

The muon reconstruction resolution is determined by comparing the reconstructed quantities with the underlying simulated information for the different muon parameters -- length and angles -- using reconstructed tracks that meet the selection criterion.

\begin{figure}[htb]
	\centering
	\includegraphics[width=0.8\textwidth]{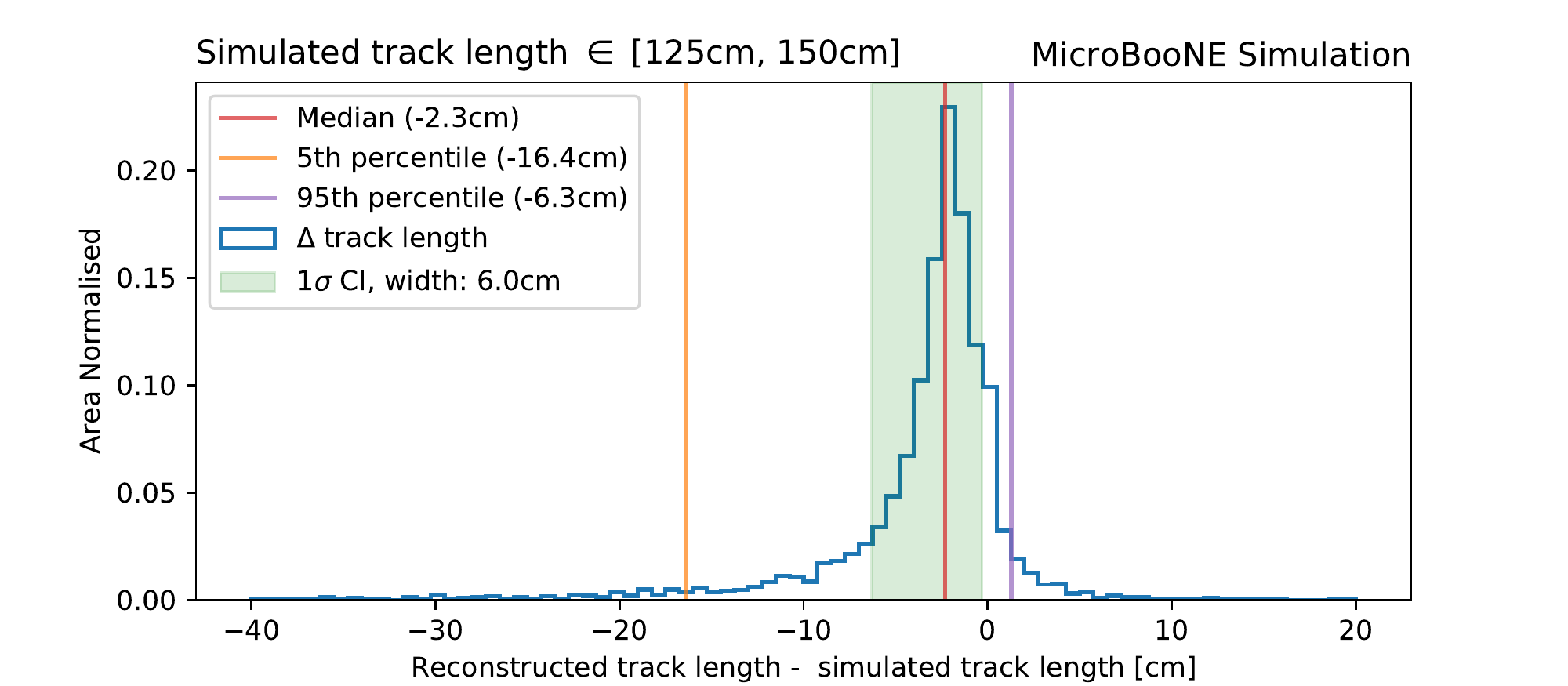}
	\caption
	{Example of how the resolution is quantified, here shown for the track length, for tracks with a simulated length between \SIlist{125;150}{\cm}. In blue, the histogram shows the difference between the simulated and reconstructed length. The median is given in red, the offset between the median and 0 is referred to as the bias. The resolution defined as half the width of the green shaded area, covering a symmetric \pct{68.3} interval around the median. The 5th (orange) and 95th (purple) percentiles define a region that includes 90\% of the tracks. The tail towards shorter reconstructed tracks is explained by tracks entering or exiting through unresponsive wire regions.}
	\label{fig:reso_example}
\end{figure}

\begin{figure}[htb]
	\centering
	\includegraphics[width=\textwidth]{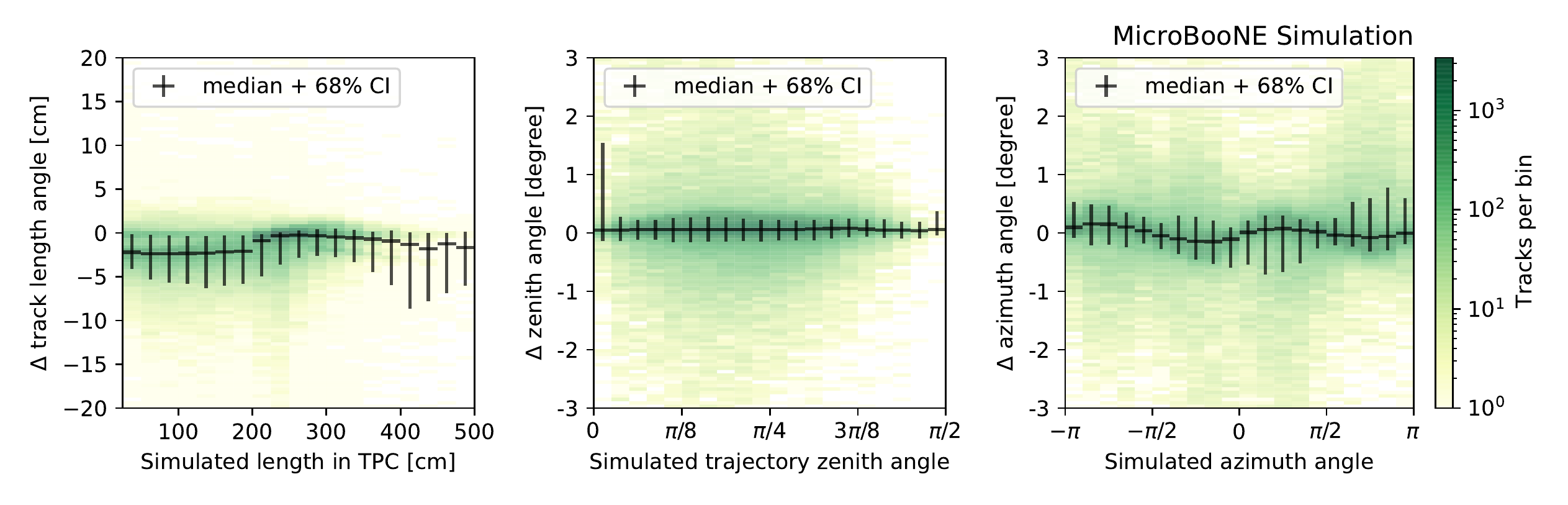}
	\caption[Reconstruction resolution for the track length, zenith angle and azimuthal angle]
	{Reconstruction resolution for the track length (left), zenith angle (middle) and azimuthal angle (right). The vertical axis shows the difference between the reconstructed and the simulated quantity. The space-charge model is applied to the simulation before comparison to data. The colour scale of the \textit{2D} histograms is logarithmic. In black, the median of each bin with the \pct{68.3} confidence interval are given.}
	\label{fig:resolution}
\end{figure}

In each bin of a simulated variable (length or angles) we find the distribution of the difference between the reconstructed and simulated variable, as illustrated in \cref{fig:reso_example}. The median is used to estimate the central value. The ranges between the median 16th and 84th percentiles respectively are used to estimate the \pct{68} confidence interval for that bin. The resolution obtained is presented in the panels of \cref{fig:resolution}.
%\begin{enumerate}
%\item In each bin of a simulated variable (length or angles) of %\cref{fig:resolution}, the difference between the reconstructed and the %simulated value is calculated for each entry.
%\item The values corresponding to the 16th, 50th and 84th percentile are %calculated in the reconstructed distribution, as shown in %\cref{fig:reso_example}. 
%\item The 50th percentile corresponds to the median and the 16th and 84th %percentiles are respectively the lower and upper bounds of the \pct{68.3} %confidence interval for every simulated variable bin. 
%\end{enumerate}

The median track length shown in the left panel of \cref{fig:resolution}, indicates that reconstructed muon segments are slightly shorter than the true length. The bias ranges from \SIrange{0.3}{2.4}{\cm}, depending on the true length. The \pct{68.3} confidence interval presents a tail towards shorter tracks, as can also be seen from the specific example in \cref{fig:reso_example}. The resolution, defined as half the width of the \pct{68.3} interval, varies from \SI{\approx1.5}{\cm}, for short and vertically crossing tracks and up to \SI{\approx 4.0}{\cm} for the longest tracks. 

The middle and right panels of \cref{fig:resolution} show the zenith and azimuth angle resolution for the muon segments. The bias of the median reconstructed angle is less than \ang{0.1} in both cases. The resolution of the zenith angle is approximately \ang{0.2}, except for the first bin which has a tail towards larger zenith angles, corresponding to tracks parallel to the collection wire plane and creating a single hit which is harder to reconstruct. The resolution of the azimuthal angle varies between \ang{0.2} and \ang{0.5}, depending on the track length and its alignment with the charge-collecting wires.

\subsection{Comparison of the Data and Predicted Rates}
\label{sc:tpc_datamc}

\begin{figure}[htb]
	\centering
	\includegraphics[width=0.75\textwidth]{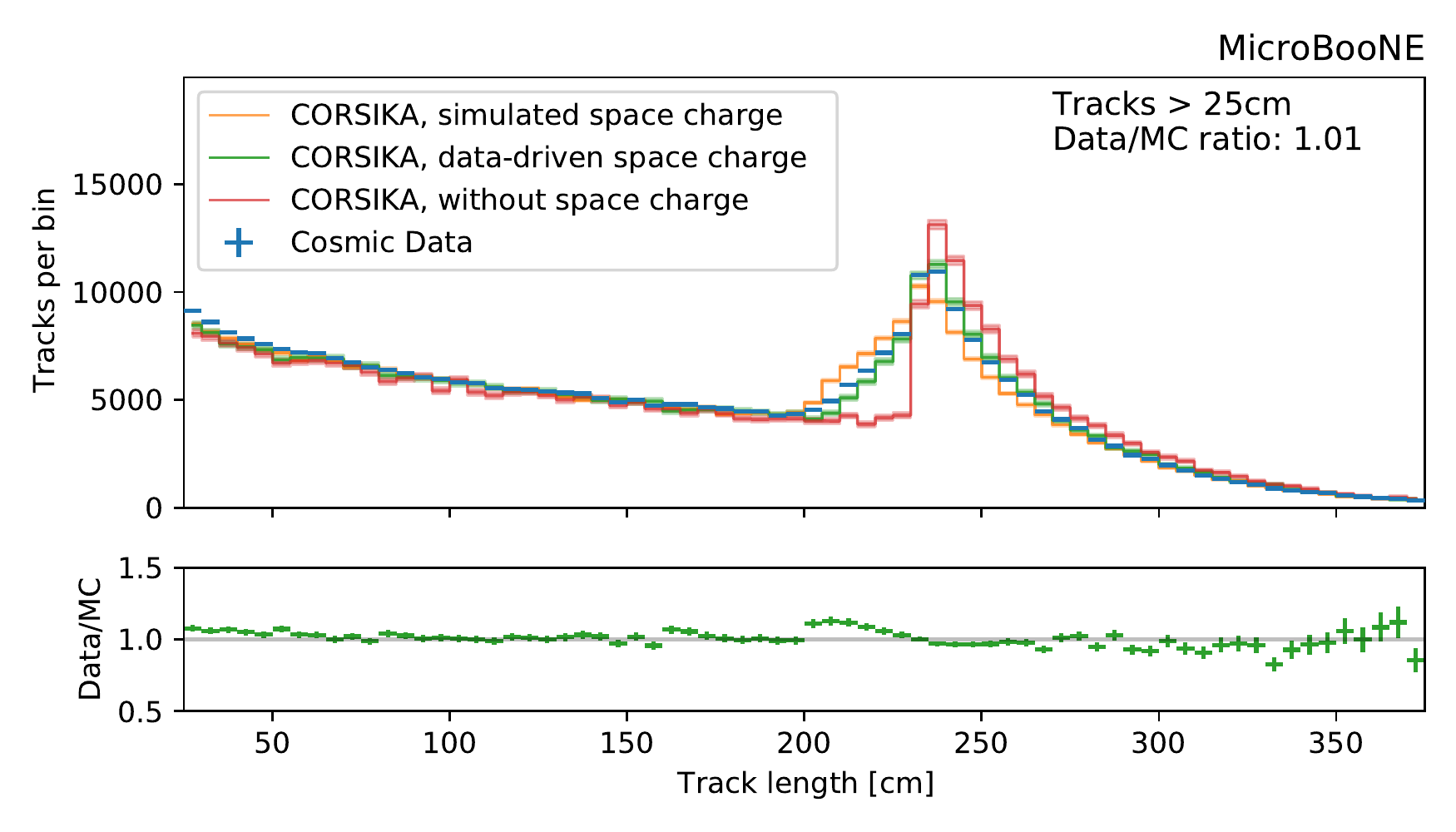}
	\caption[Reconstructed atmospheric muon track length]
	{Comparison of the reconstructed track length in data (blue crosses) and \cor Monte Carlo simulation,  with data-driven (green), with theory-based (orange), and with no (red) space charge effect included. The bin width on the horizontal axis is \SI{5}{\cm}. The bottom panel shows the ratio of data to simulation in green (Data/MC) for the sample with data-driven space charge. The overall ratio is \num{1.014+-0.004} (stat).}
	\label{fig:tracklength}
\end{figure}

\begin{figure}[htb]
	\centering
	\includegraphics[width=0.75\textwidth]{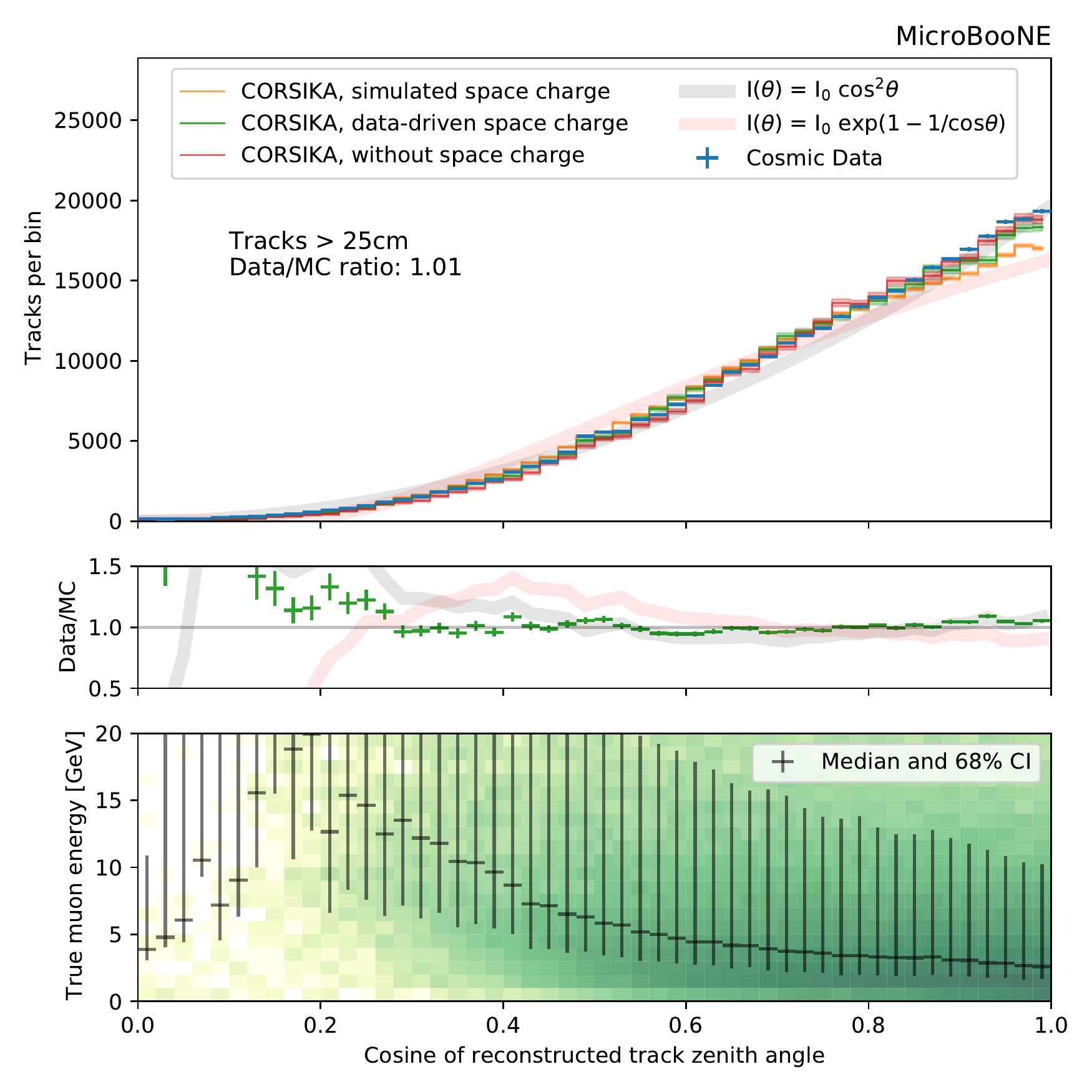}
	\caption[Reconstructed atmospheric muon zenith angle]
	{Comparison of the cosine of the reconstructed zenith angle in data (blue crosses) and \cor Monte Carlo simulation, with data-driven (green), with theory-based (orange), and with no (red) space charge effect included. The middle panel shows the ratio of data to simulation for the sample with data-driven space charge. The shaded bands (grey and red) are area normalised analytical models of the zenith dependence of atmospheric muons. In the middle panel, those models are compared with the data-driven space charge sample.	The \textit{2D} histogram in the bottom panel shows the true muon energy for the data-driven space charge sample. For each bin, the horizontal black line indicates the median true energy and the bars correspond to the \pct{68.3} interval. Although the full phase space of downward-going tracks is given, the simulation significantly undercovers the horizontally oriented tracks -- \ang{75} and higher -- due to the finite extent of the simulation. See \cref{sc:tpc_syst} for a further discussion. }
	\label{fig:trackzenith}
\end{figure}

\begin{figure}[htb]
	\centering
	\includegraphics[width=0.75\textwidth]{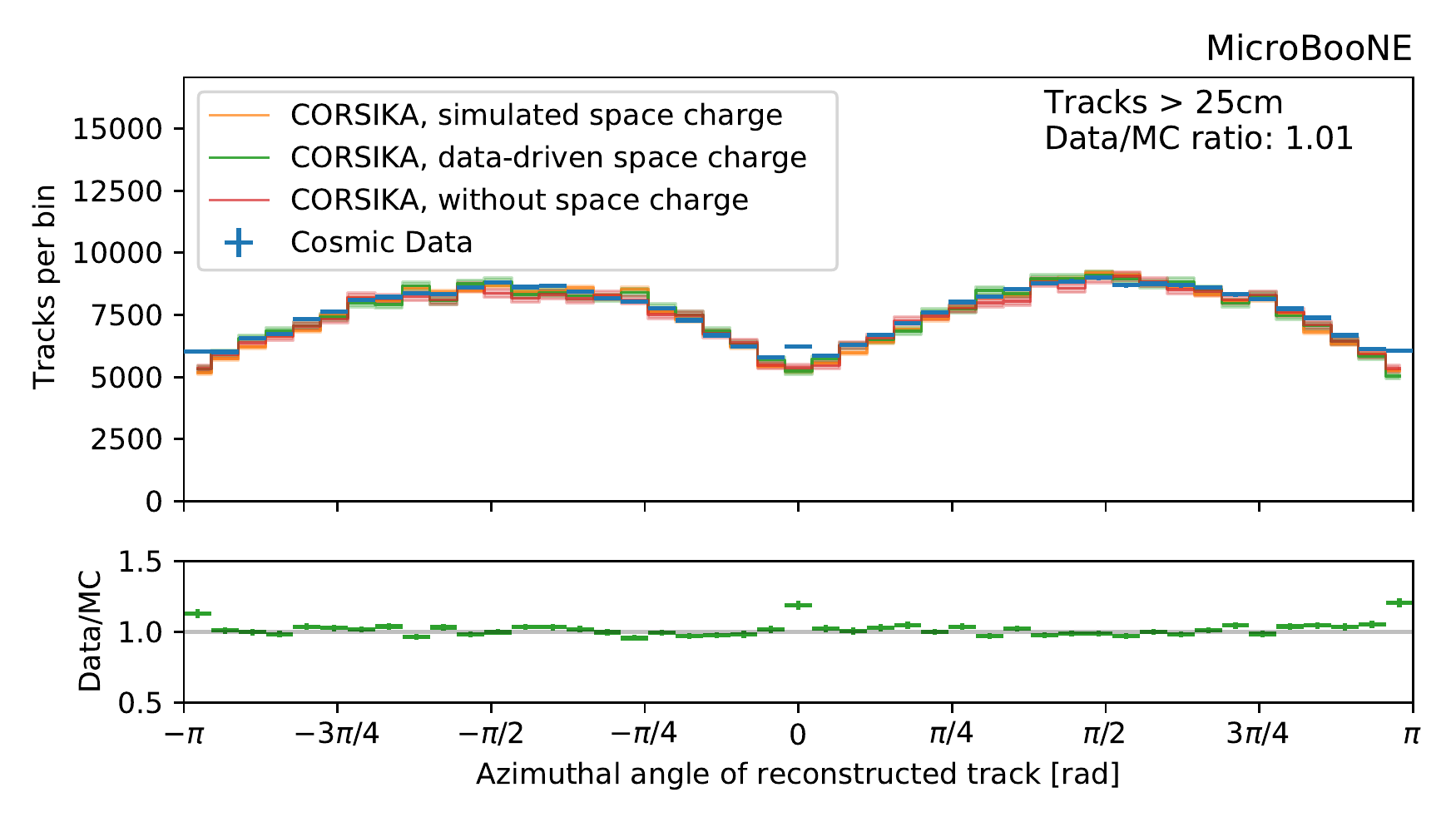}
	\caption[Reconstructed atmospheric muon azimuth angle]
	{Comparison of the reconstructed track azimuth angle in data (blue crosses) and \cor Monte Carlo, with data-driven (green), with theory-based (orange), and with no (red) space charge effect included. The bottom panel shows the ratio of data to simulation for the sample with data-driven space charge. The bin width on the horizontal axis is $8^\circ$}
	\label{fig:trackazimuth}
\end{figure}

In \cref{fig:tracklength,fig:trackazimuth,fig:trackzenith}, events from the unbiased data stream are compared with simulated cosmic events. The data sample was collected between February and May 2016 and comprises 25k events, each consisting of a TPC readout window of \SI{3.2}{\ms}. The sample has no overlap with accelerator neutrino triggers.

As introduced in \cref{ss:sce}, two variants of the space charge effect model are used in the simulated samples. A data-driven approach is used as the nominal model, indicated as \textit{data-driven space charge} and shown in green. The theoretically derived correction is referred to as \textit{simulated space charge} and shown in orange. A case where no space charge effect is  taken into account, demonstrating the magnitude and shape of its impact, is also included and shown in red. The data-driven space charge correction is taken as the default (central value) and the corresponding ratio of data to simulation, integrated over the muon momentum and direction, is \num{1.014+-0.004} (stat). The \systs will be discussed in \cref{sc:tpc_syst}.

The track length distribution in \cref{fig:tracklength} is peaked at \SI{\approx 2.3}{\m}, corresponding to the full height of the TPC, where top-bottom through-going muons pile up. A sharp turn on of the peak can be seen when no space charge is simulated (red). When the space charge effect is taken into account, the peak gets smeared out towards shorter reconstructed track lengths. The data in this region is situated between the two different space charge models, indicating that the sample with theory-based space charge correction (green) slightly exaggerates the effect and the sample with the data-driven model (orange) slightly underestimates it.

The comparison of the cosine of the reconstructed zenith angle in data and simulation is shown in \cref{fig:trackzenith}. The zenith angle has an intrinsic energy dependence due to the centre of the TPC being located \SI{6}{\m} underground. Muons at larger zenith angles, being more horizontal, traverse a larger distance in both the atmosphere and the dirt surrounding LArTF before entering the cryostat. This is studied in the bottom panel of \cref{fig:trackzenith}. The bulk of the muons entering the TPC have an energy below \SI{3}{\GeV}. For muons entering more horizontally, the median energy increases up to \SI{15}{\GeV}. Due to the limited extent of the simulation, zenith angles above \ang{75} are not well covered and the simulation underpredicts the horizontal muon flux. 

The zenith angle dependence is compared to two analytical models~\cite{grieder}. In the specific case of these two analytical models, the models are normalised to the data to compare the shapes, as shown by the shaded bands in the top and middle panel of \cref{fig:trackzenith}. These models do not account for the detector acceptance and shielding; nevertheless, their shape follows the simulation and data. 
%At higher zenith angles, the discrepancy is expected due to the significant amount of ground the tracks need to traverse before reaching the detector.

The generation of atmospheric muons is isotropic in the azimuthal space, but the acceptance of the cuboid shape of the \uB TPC affects the observed distribution, as can be seen in \cref{fig:trackazimuth}. Small features in the ratio of data to simulation at \ang{0} and \ang{180} are caused by tracks parallel to the wire planes, where the drifted charge over the whole track length arrives coincidentally, making both the reconstruction and modelling challenging. Removal of coherent noise, in particular, affects the reconstruction for this topology~\cite{ub_noise}.

\subsection{Systematic Uncertainty Estimation}
\label{sc:tpc_syst}

\begin{figure}[htb]
	\centering
	\includegraphics[width=0.85\textwidth]{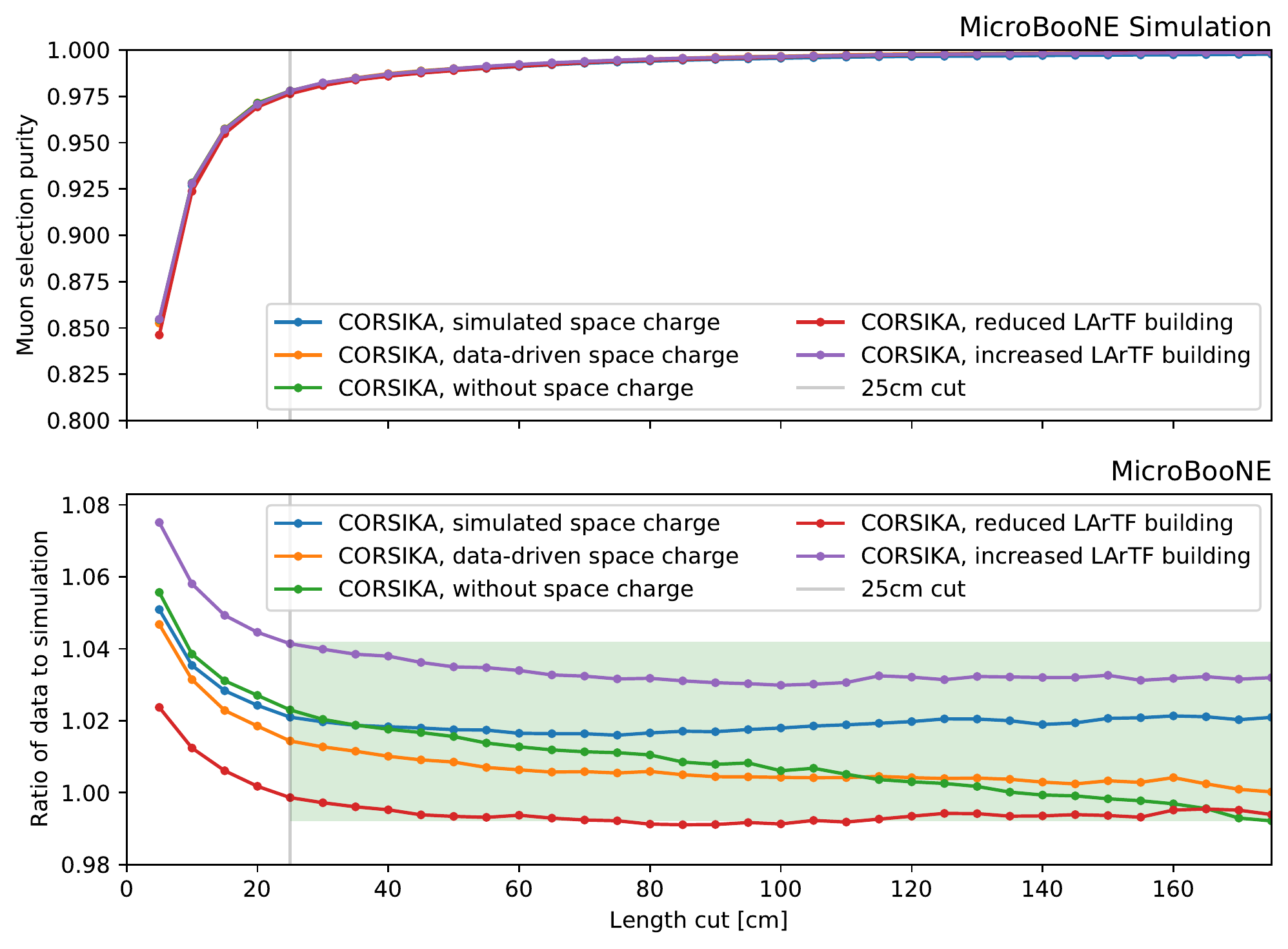}
	\caption[Effect of systematic uncertainties on the TPC-based atmospheric muon measurement]
	{Top: Impact of different length criteria on the muon selection purity. Bottom: The ratio of the data to simulation rates as a function of the minimum track length requirement, for different simulated samples. In orange, the purity (top) and ratio (bottom) are given using the sample with data-driven space charge effect included. In blue, a sample with the theory-based space charge simulation is shown. In green, a sample without simulated space charge is shown. The purple and red curves correspond to samples with data-driven space charge effect where the concrete density of the \lartf building is varied with \pct{\pm 30}. The green band represents the combined \syst as obtained in \cref{tab:tpc_syst}. The grey line shows the \SI{25}{\cm} requirement used in this work.}
	\label{fig:syst}
\end{figure}

Using the central value simulation, as introduced in \cref{ch:sim} and including a data-driven space charge effect correction, the ratio of data to simulation is \num{1.014+-0.004} (stat). Four sources of \systs that impact the ratio are evaluated:
\begin{enumerate}

\item \textit{Incomplete angular coverage.} Limitations of the simulation show up as discrepancies between the data and simulation in some regions of the distributions in \cref{fig:trackzenith,fig:trackazimuth}. The difference between the data and the simulation calculated with and without these regions is taken as a \syst.
\begin{itemize}
\item \textit{Zenith angle.} For zenith angles between \ang{3} and \ang{70}, the ratio is \num{1.010}.
\item \textit{Azimuth angle.} Without a region of \ang{2} on either side of the azimuthal \ang{0} and \ang{180} -- where the tracks are parallel to the collection plane -- the ratio is \num{1.008}.
\end{itemize}

\item \textit{Length criterion.} An increased muon purity can be achieved by increasing the track length criterion above \SI{25}{\cm}. In the top panel of \cref{fig:syst}, the purity is evaluated for minimum lengths ranging from \SIrange{5}{175}{\cm}. In the bottom panel, the effect on the ratio of data to simulation is shown. For the simulated sample with the data-driven space charge effect, the ratio changes from 1.014 to 1.000 when changing the track length requirement from \SIrange{25}{175}{\cm}.

\item \textit{Space charge effect.} As demonstrated in \cref{fig:tracklength}, the space-charge effect shortens the reconstructed length of tracks. In the bottom panel of \cref{fig:syst}, the ratio of data to simulation is evaluated for different space charge models. The difference between the theory and data-driven model at a length criterion of \SI{25}{\cm} is \pct{0.7}.

\item \textit{\lartf building geometry.} The details of the geometry used for the simulations, such as the concrete walls or the roof can impact the simulated atmospheric muon rate in the TPC. The potential effect is conservatively estimated by varying the concrete density of the walls and roof by \pct{+30} (increased \lartf building) and \pct{-30} (reduced \lartf building), leading to an increase of \pct{2.7} and decrease of \pct{1.5} in the ratio, respectively.
\end{enumerate} 
The different sources of \syst are listed in \cref{tab:tpc_syst} and are combined in quadrature, resulting in a ratio of data to simulation of
\begin{center}
\num{1.014+-0.004} (stat)$\: \substack{+0.028 \\ -0.022}\:$(sys)
\end{center}
The size of the \syst is indicated by the green shaded area in the bottom panel of \cref{fig:syst} and covers all the values of the ratio obtained from combining different samples and different length criteria. 

\begin{table}[htb]
\centering
\caption{Contribution of the different systematic uncertainties on the TPC muon rate measurement.}
\vspace{3mm}
\label{tab:tpc_syst}
\begin{tabular}{lc}
\toprule
Systematic variation      & Uncertainty \\ \midrule
Zenith angle phase space  & \pct{-0.4}            \\
Azimuth angle phase space & \pct{-0.6}            \\
Length criterion          & \pct{-1.4}            \\
Space charge model        & \pct{+0.7}           \\
Increased LArTF shielding & \pct{+2.7}            \\
Decreased LArTF shielding & \pct{-1.5}            \\ 
Total                  & \pct{-2.2} / \pct{+2.8} \\ \bottomrule
\end{tabular}
\end{table}

\section{Muon Rate Measurement using the Photon Detection System}
\label{ch:pmt}

The muon rate is independently measured using \uB 's optical system, consisting of 32 8-inch PMTs. The rate of optically reconstructed signals can be related to the atmospheric muon-rate, validating the TPC-based measurement. 

\subsection{Optical Reconstruction}
\label{sc:pmt_rec}

%  PEThreshold:    30
%  MinPECoinc:     8
%  MinMultCoinc:   2
%  IntegralTime:   0.625
%  PreSample:      0.09375
%  VetoSize:       8.

\begin{figure}[htb]
\centering
\includegraphics[width=\textwidth]{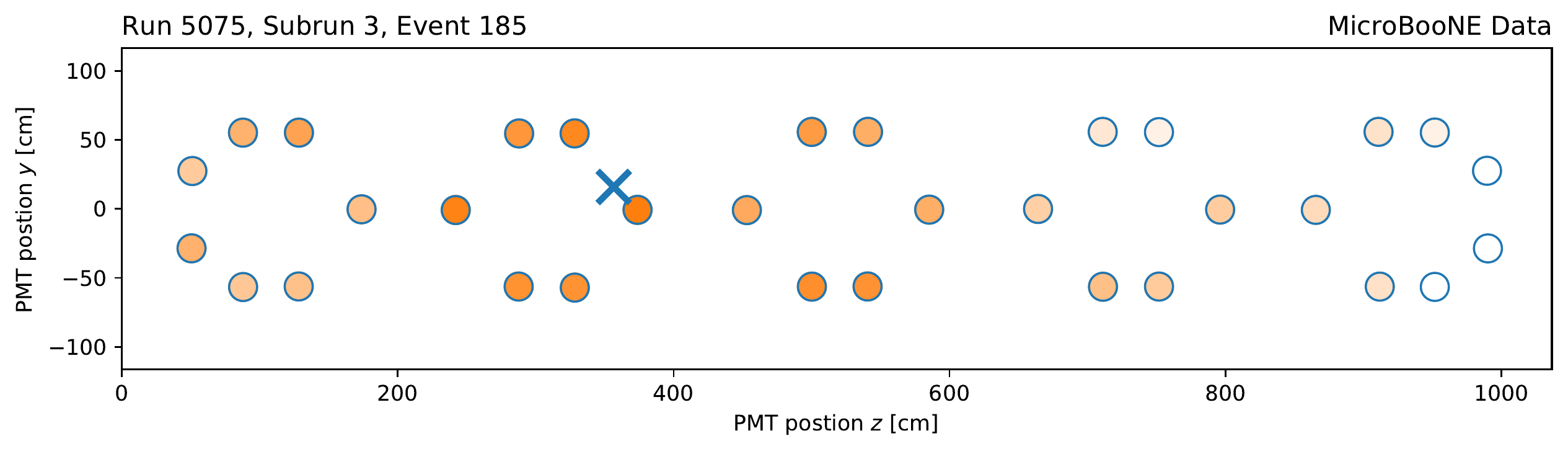}
\caption[Example of a flash consisting of multiple PMT hits combined]{Example of a flash, consisting of multiple PMT signals combined. The orange colour scale is logarithmic and represents the intensity of the light, arriving in each PMT, integrated over \SI{625}{\ns}. The blue cross indicates the location of the centre of the flash. The 32 PMTs are organised in five rosettes, located behind the wires in the $yz$-plane.}
\label{fig:flash}
\end{figure}

The optical reconstruction builds flashes from the PMT waveforms. Flashes represent coincidental optical activity across several PMTs, usually caused by a particle interaction in the TPC. An example is illustrated in \cref{fig:flash}. A flash object is created when at least two PMTs detect a signal corresponding to at least 8 photoelectrons (PEs), each in a coincidence interval of \SI{94}{\ns}. The flash object contains the light of the different PMTs that crossed the \SI{8}{PE} threshold, integrated during \SI{625}{\ns}. This window covers the \SI{6}{\ns} prompt (and a fraction of the \SI{1.5}{\us} slow) scintillation component of liquid argon. A veto window of \SI{8}{\us} is applied after the creation of a flash to avoid the creation of spurious flashes due to the tail of the slow scintillation light component. %A more detailed description of the optical system and \textit{flash} reconstruction can be found in~\cite{ub_pmt_note}.

\subsection{Optical Interaction Rate}
\label{sc:pmt_rate}

\paragraph{Purity} In simulation, \pct{94.1\pm0.1} (stat) of the reconstructed flashes are matched to simulated muons. The majority of the remaining fraction originated from light induced by inelastic neutron scattering. The \pct{94.1\pm0.1} (stat) can be divided into muons entering the TPC (\pct{81.6\pm0.1}), and flashes caused by muons that enter the cylindrical cryostat but do not pass through the cuboid TPC volume \pct{12.6\pm0.1} (stat).

\begin{figure}[htb]
	\centering
	\includegraphics[height=4.5cm]{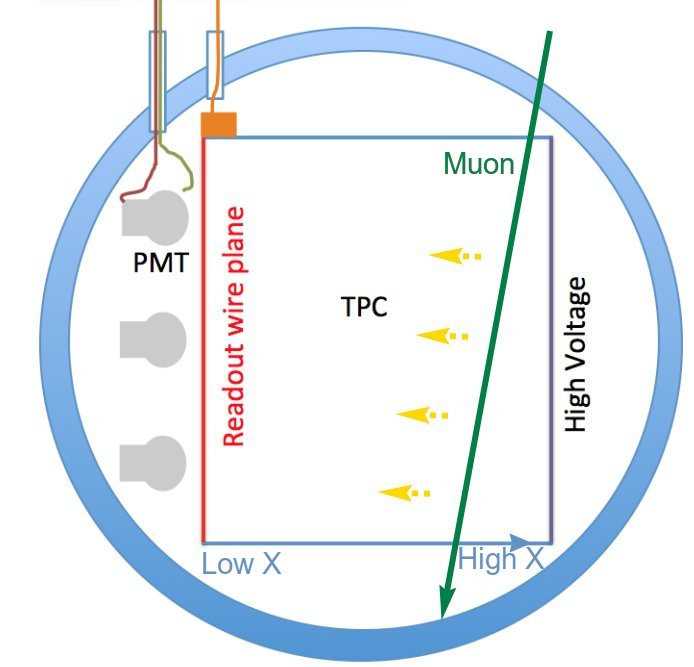}\hfill
	\includegraphics[height=4.5cm]{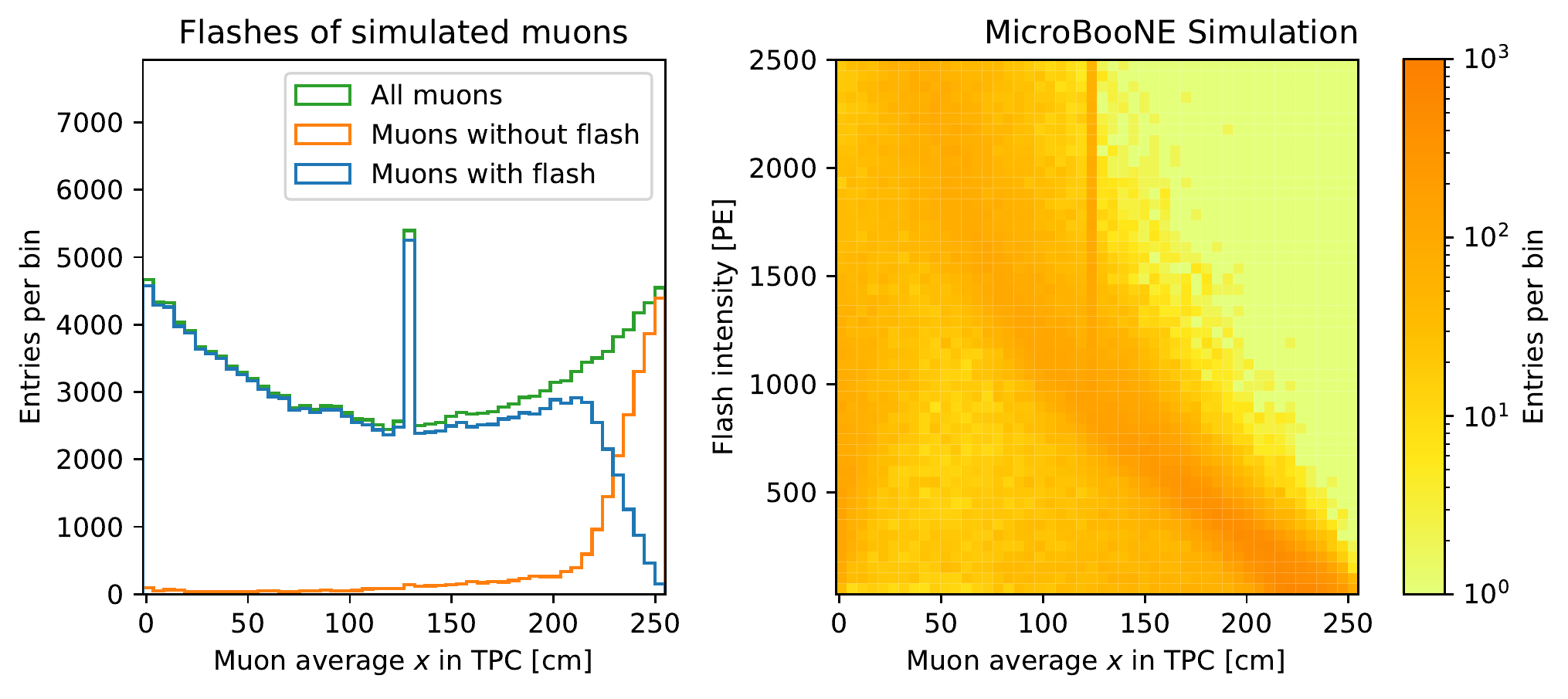}
	\caption[Flash reconstruction efficiency for comic muon as a function of $x$]
	{(Left) Schematic side view of the \uB cryostat, showing a muon crossing near the cathode (high $x$). (Middle) Average $x$ position of the crossing muon trajectory inside the active TPC volume. The simulated muons (green) are divided in those that have a corresponding reconstructed flash \pct{81.6\pm0.1} (blue), and those that do not (orange). The central peak corresponds to the average $x$ location of anode-cathode crossing muons. Note the sharp decline for tracks situated within half a meter from the cathode plane. (Right) The flash intensity versus the average muon distance from the wire planes. The logarithmic colour scale represents the number of flashes per bin. Far from the anode, the flash intensity is strongly reduced.}
	\label{fig:pmt_eff}
\end{figure}

\paragraph{Reconstruction efficiency} Of the simulated atmospheric muons entering the TPC, \pct{81.3\pm0.3} (stat) lead to a reconstructed flash. The \SI{8}{\us} veto window causes an additional loss of \pct{4.8} due to dead time. A further inefficiency is due to geometrical effects, illustrated in \cref{fig:pmt_eff}, where the flash efficiency is reduced for simulated muons at high $x$, far from the PMTs. A similar, although less pronounced, effect is found at the TPC edges in the $yz$-plane and is accounted for as a systematic uncertainty in \cref{sc:pmt_syst}. 

After correcting for the dead time introduced by the veto window, the flash rate in simulation is \SI{4.642 +- 0.011}{\kHz} (stat). Using the same data sample as in \cref{sc:tpc_datamc}, a flash rate of \SI{4.593 +- 0.007}{\kHz} (stat) is obtained, after the dead-time correction. These results do not include \systs, which are described in the next section.
% Considering only the statistical uncertainty, an excess in simulation over data of \pct{1.0+-0.3} is observed.

\subsection{Systematic Uncertainty Estimation}
\label{sc:pmt_syst}

The sources of systematic error can be divided into light modelling systematic uncertainties and the uncertainty from the \lartf geometry simulation and are listed below: 

\begin{enumerate}
    \item \textit{Flash photoelectron threshold.} Flashes are created with a minimum of optical-activity of \SI{34}{PE}. This threshold can be increased to evaluate the sensitivity of the ratio between data and simulation due to the very dimmest flashes. After doubling the threshold to \SI{68}{PE} in both data and simulation, this ratio changes from \numrange{0.990}{0.992}, resulting in a systematic uncertainty of \pct{0.2}. The effect of this change is shown in the right panel of \cref{fig:flash_pe}.
    
    \begin{figure}
        \centering
        \includegraphics[width=0.8\textwidth]{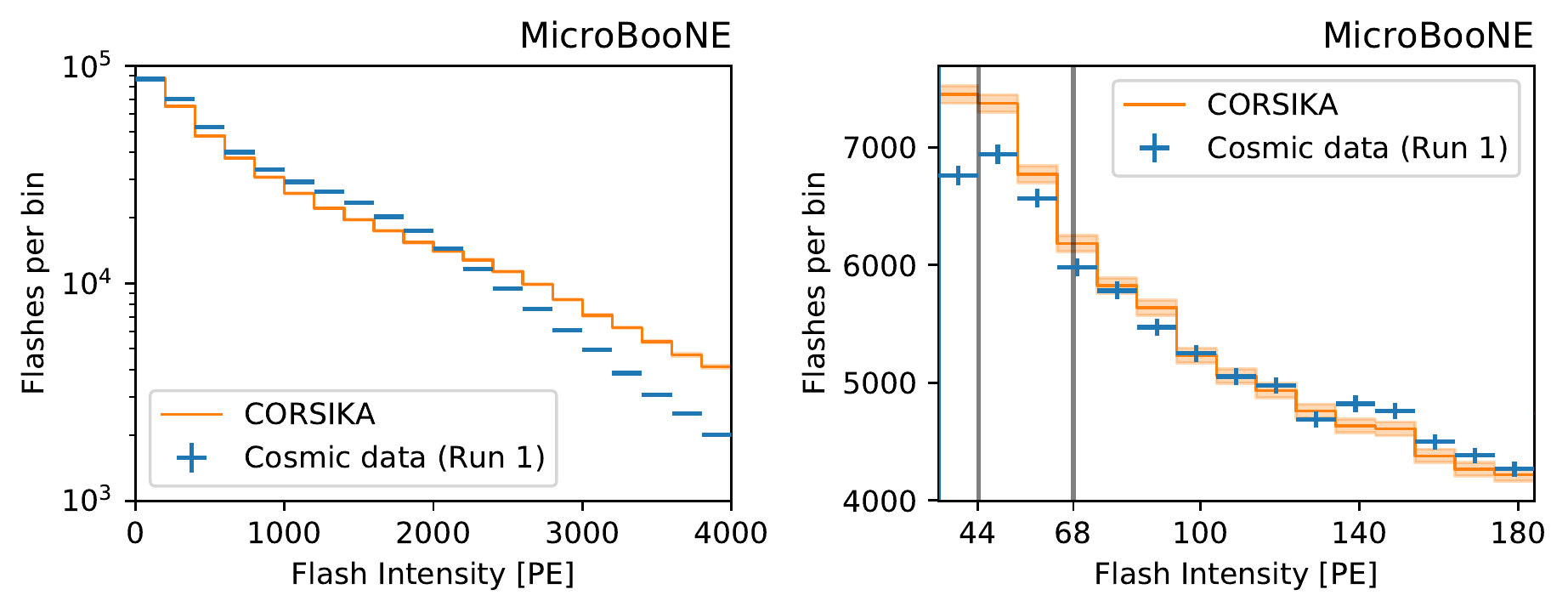}
        \caption{Comparison of the flash intensity, expressed as a number of photoelectrons (PE), in simulation and data. The right panel zooms in for low-intensity flashes. The two vertical grey lines correspond to the investigation of systematic uncertainties introduced by the increased threshold (\SI{68}{PE}) and the \pct{\pm30} variation in light yield (\SI{44.2}{PE}). The shaded area (CORSIKA Simulation) and vertical crosses (Cosmic Data) represent the statistical variation.}
        \label{fig:flash_pe}
    \end{figure}
    
    \item \textit{Light yield variations.} The impact of a difference in absolute light yield between simulation and data is evaluated by varying the total flash intensity by \pct{\pm30} in simulation. 
    The change in light yield manifests itself as a different threshold in data and simulation, as illustrated in the right panel of \cref{fig:flash_pe}. The decreased scintillation light production leads to a data to simulation ratio of \num{1.006}, the increased light yield leads to a ratio of \num{0.974}. Therefore this uncertainty is symmetric and affects the result by \pct{1.6}. 
    
    \item \textit{Out-of-TPC light modelling.} As discussed in \cref{sc:pmt_rate}, a fraction of the flashes are caused by muons crossing the cryostat without entering the TPC. Their impact on the ratio can be studied by restricting the measurement to flashes which are centred inside a constrained region in the $yz$-plane (green shaded area in \cref{fig:pmt_syst}). 
    Excluding flashes that are centred less then \SI{1}{\m} away from the edges in the $z$-direction and less than \SI{30}{\cm} from the top in the vertical direction, the contribution of flashes caused by muons inside the cryostat without entering the TPC decreases from \pct{12.6} to \pct{6.2}. The ratio of data to simulation changes from \num{0.990} to \num{0.992}. The \syst is taken to be twice the difference in the ratio to account for the \pct{6.2} remaining contribution of flashes caused by activity outside of the TPC.
    
    \begin{figure}[htb]
	\centering
	\includegraphics[width=\textwidth]{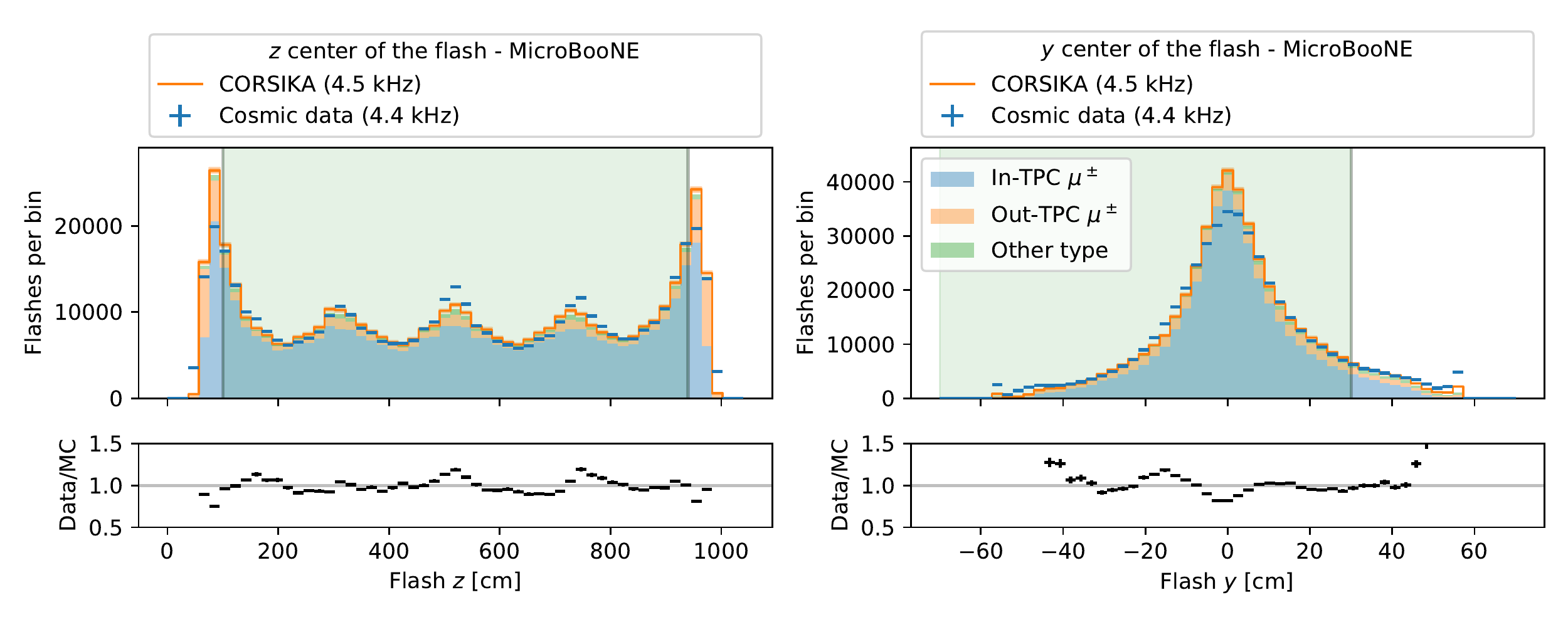}
	\caption[Centre of the reconstructed flash for cosmogenic muons]
	{Position of the centre of the flash in the $z$ (left) and $y$ (right) directions. The flashes from muons entering the TPC are shown in blue, from muons entering the cryostat but not the TPC in orange, and a small contribution of non-muon particles is shown in green. The shaded green indicates the restricted area where the contribution of out-TPC muons is reduced. The five bumps structure is explained by the position of the PMTs as shown in \cref{fig:flash}.}
	\label{fig:pmt_syst}
    \end{figure}

    \item \textit{\lartf geometry modelling} To account for potential mismodeling of the building geometry, two samples, one with increased and one with decreased concrete density, are used to estimate the effect on the flash rate. Increasing the concrete density by \pct{30} leads to a reduction in flash-rate of \pct{2.7}, while decreasing the density with \pct{30} increases the flash-rate by \pct{2.3}.
\end{enumerate}

The different sources of \syst are listed in \cref{tab:pmt_syst}, the total systematic error is obtained by adding the different contributions in quadrature. The ratio of data to simulation is:
\begin{center}
\num{0.990+-0.003} (stat)$\: \substack{+0.032 \\ -0.028}\:$(sys).
\end{center}

\begin{table}[htb]
\centering
\caption{Systematic uncertainties on the PMT muon rate measurement.}
\vspace{3mm}
\label{tab:pmt_syst}
\begin{tabular}{lc}
\toprule
Systematic variation      & Uncertainty \\ \midrule
Flash photoelectron threshold  & \pct{+0.2}       \\
Light yield variations          & \pct{\pm 1.6}                             \\
Out-of-TPC light modelling      & \pct{+0.4}       \\
Increased LArTF shielding & \pct{+2.7}             \\
Decreased LArTF shielding & \pct{-2.3}                                      \\ 
Total                  & \pct{-2.8} / \pct{+3.2} \\ \bottomrule
\end{tabular}
\end{table}

\section{Conclusion}
\label{ch:res}

In \cref{ch:tpc,ch:pmt}, we present ratios between the muon rate in data and simulation. These ratios can be converted into the muon flux per area at the position of the detector. To obtain this rate, the simulated rate is scaled by the ratio between the data and simulated rate. This rate does not depend on the systematic uncertainty introduced by the \lartf geometry since the variation in simulated shielding cancels out in the observed ratio. The atmospheric muon rate obtained independently with the TPC and the PMTs is quoted with systematic uncertainty in the first column of \cref{tab:results}. Excluding the systematic uncertainty related to the \lartf geometry, the two measurements and their \systs are uncorrelated. All rates are in agreement within errors. No systematic uncertainties due to the choice of \cor configuration, as was introduced in \cref{sec:model}, is included in these results. 

Relying on the simulation of the \lartf building geometry, the measured ratios are extrapolated to the muon rate at the Earth's surface, which is useful for near-surface experiments located at Fermilab. The obtained atmospheric muon flux is compared with the \cor rates in the last column of \cref{tab:results}. In this case, the dominant systematic uncertainty originates from the \lartf building geometry modelling and is strongly correlated between the two measurements. These measurements include muons with a momentum $\gtrsim\SI{0.3}{\GeV \per c}$ (see \cref{ss:sim_rate}), but the bulk of the detected muons are in the \SIrange{1}{3}{\GeV \per c} momentum range, as can be seen in the bottom panel of \cref{fig:trackzenith}.

\begin{figure}[htb]
	\centering
	\includegraphics[width=0.75\textwidth]{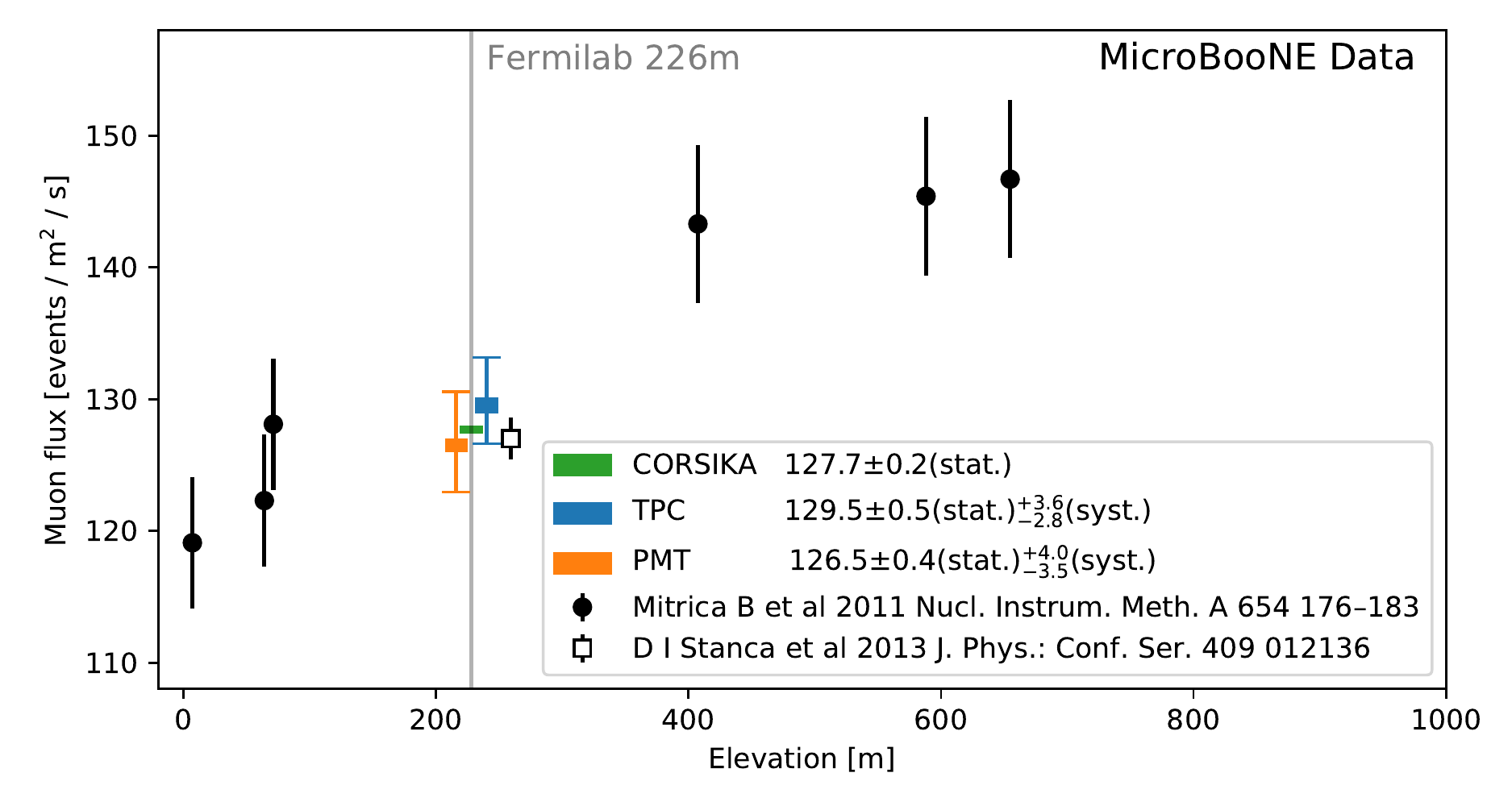}
	\caption[Comparison of the muon flux measurement in this work with previous work]
	{Comparison of the integrated muon flux measurement from this work (coloured data points) with previous measurements~\cite{muon,muon2} (black data points), for different elevations. The vertical grey line indicates the elevation of Fermilab. The TPC (blue) and PMT (orange) measurements are compared with the \cor prediction (green). The vertical width of the data points represents the statistical error, and the error bars represent the combined statistical and systematic errors added in quadrature. The blue and orange points are horizontally offset from the grey line for visual clarity.}
	\label{fig:result}
\end{figure}

The integrated muon flux obtained with the \uB detector is compared with other measurements in \cref{fig:result}. The results presented are in agreement with both the \cor prediction of \SI{127.7+-0.2}{\Hzz \per \m^2} and previous measurements~\cite{previous,muon,muon2}. Because of the choice to integrate over the angular and muon momentum space, the result can be compared with differential measurements using \cref{eq:int_intensity}. 

The data used in this works spans a period of four months (February to May 2016), during which no significant evidence of seasonal fluctuations was observed. The measured tracks are dominated by low-energetic muons due to the lack of shielding. Furthermore, the detector technology is unable to discriminate tracks based on the muon momentum. In these conditions, any seasonal effects are expected to be challenging to detect at MicroBooNE~\cite{seasonal}.

\begin{table}[htb]
\small
\centering
\caption[Comparison of the measured and simulated atmospheric muon rates in \uB]
{Summary of the atmospheric muon rates in simulation and data, at the TPC and extrapolated to \lartf building roof level. The simulation rates using \cor is presented, along with the measured TPC and PMT data rates, as obtained in \cref{ch:tpc,ch:pmt} respectively.}
\vspace{3mm}
\label{tab:results}
\begin{tabular}{@{}llll@{}}
\toprule
                               &                                         & Rate at TPC (\si{\Hzz \per \m^2})  
                                                                         & \multicolumn{1}{c}{Rate above roof (\si{\Hzz \per \m^2})}     \\ \midrule
                                                              
Simulation & \cor default              & \num{111.6+-0.3}                                
                                                                         & \num{127.7+-0.2}                                              \\
 %                              & \cor + CMC     & \num{140.1+-0.3}                        
 %                                                                        & \num{160.9+-0.3}                                              \\
                               &                                         &                                         &                     \\
\multirow{2}{*}{Data}          & TPC   \vspace{1mm}           & \num{113.2+-0.5} (stat)$\: \substack{+0.8 \\ -1.8}\:$(sys)          
                                                                         & \num{129.5+-0.5} (stat)$\: \substack{+3.6 \\ -2.8}\:$(sys) \\
                               & PMT                          & \num{110.4+-0.4} (stat)$\: \substack{+1.8 \\ -2.0}\:$(sys) 
                                                                         & \num{126.5+-0.4} (stat)$\: \substack{+4.1 \\ -3.5}\:$(sys) \\ \bottomrule
\end{tabular}
\end{table}

\acknowledgments
This document is prepared by the MicroBooNE collaboration using the resources of the Fermi National Accelerator Laboratory (Fermilab), a U.S. Department of Energy, Office of Science, HEP User Facility. Fermilab is managed by Fermi Research Alliance, LLC (FRA), acting under Contract
No. DE-AC02-07CH11359. MicroBooNE is supported by the following: the U.S. Department of Energy, Office of Science, Offices of High Energy Physics and Nuclear Physics; the U.S. National Science Foundation; the Swiss National Science Foundation; the Science and Technology Facilities Council of the United Kingdom; and The Royal Society (United Kingdom). This research is further supported by the U.S. Department of Energy, Office of Science, Office of High Energy Physics, under the Award Number DE-SC0007881. Additional support was received from Oxford University Press and Jesus College Oxford.

%\paragraph{Note added.} This is also a good position for notes added
%after the paper has been written.
% We suggest to always provide author, title and journal data:
\bibliographystyle{JHEP}
\bibliography{biblio}

\providecommand{\href}[2]{#2}\begingroup\raggedright\begin{thebibliography}{10}

\bibitem{det}
{\scshape MicroBooNE} collaboration, \emph{Design and construction of the
  {MicroBooNE} detector},
  \href{https://doi.org/10.1088/1748-0221/12/02/P02017}{\emph{JINST} {\bfseries
  12} (2017) P02017} [\href{https://arxiv.org/abs/1612.05824}{{\ttfamily
  1612.05824}}].

\bibitem{mu2e}
{\scshape Mu2e} collaboration, \emph{{The Mu2e Experiment}},
  \href{https://doi.org/10.3389/fphy.2019.00001}{\emph{Front. in Phys.}
  {\bfseries 7} (2019) 1} [\href{https://arxiv.org/abs/1901.11099}{{\ttfamily
  1901.11099}}].

\bibitem{sbn}
P.~A. Machado, O.~Palamara and D.~W. Schmitz, \emph{The short-baseline neutrino
  program at {Fermilab}},
  \href{https://doi.org/10.1146/annurev-nucl-101917-020949}{\emph{Annu. Rev.
  Nucl. Part. Sci.} {\bfseries 69} (2019) 363}
  [\href{https://arxiv.org/abs/1903.04608}{{\ttfamily 1903.04608}}].

\bibitem{dunend}
{\scshape DUNE} collaboration, \emph{Deep underground neutrino experiment
  ({DUNE}), far detector technical design report, volume {I} introduction to
  {DUNE}}, \href{https://doi.org/10.1088/1748-0221/15/08/T08008}{\emph{JINST}
  {\bfseries 15} (2020) T08008}
  [\href{https://arxiv.org/abs/2002.02967}{{\ttfamily 2002.02967}}].

\bibitem{corsika}
D.~Heck, J.~Knapp, J.~Capdevielle, G.~Schatz and T.~Thouw, \emph{{CORSIKA}: a
  {Monte Carlo} code to simulate extensive air showers}. KIT, 1998,
  \href{https://doi.org/10.5445/IR/270043064}{10.5445/IR/270043064}.

\bibitem{cry}
C.~Hagmann, D.~Lange and D.~Wright, \emph{Cosmic-ray shower generator ({CRY})
  for {Monte Carlo} transport codes},
  \href{https://doi.org/10.1109/NSSMIC.2007.4437209}{\emph{IEEE Nucl. Sci.
  Symp. Conf. Rec.} {\bfseries 2} (2007) 1143 }.

\bibitem{comptongetting}
A.~H. Compton and I.~A. Getting, \emph{An apparent effect of galactic rotation
  on the intensity of cosmic rays},
  \href{https://doi.org/10.1103/PhysRev.47.817}{\emph{Phys. Rev.} {\bfseries
  47} (1935) 817}.

\bibitem{pdg}
{P.A. Zyla et al. (Particle Data Group)}, \emph{{Review of Particle Physics}},
  \href{https://doi.org/10.1093/ptep/ptaa104}{\emph{Prog. Theor. Exp. Phys.}
  {\bfseries 2020} (2020) 083C01}.

\bibitem{pamela}
{\scshape PAMELA} collaboration, \emph{Measurements of cosmic-ray proton and
  helium spectra},
  \href{https://doi.org/10.1126/science.1199172}{\emph{Science} {\bfseries 332}
  (2011) 69} [\href{https://arxiv.org/abs/1103.4055}{{\ttfamily 1103.4055}}].

\bibitem{primarydatabase}
D.~{Maurin}, F.~{Melot} and R.~{Taillet}, \emph{A database of charged cosmic
  rays}, \href{https://doi.org/10.1051/0004-6361/201321344}{\emph{A\&A}
  {\bfseries 569} (2014) A32}
  [\href{https://arxiv.org/abs/1302.5525}{{\ttfamily 1302.5525}}].

\bibitem{1990}
C.~Forti, H.~Bilokon, B.~d'Ettorre Piazzoli, T.~K. Gaisser, L.~Satta and
  T.~Stanev, \emph{Simulation of atmospheric cascades and deep-underground
  muons}, \href{https://doi.org/10.1103/PhysRevD.42.3668}{\emph{Phys. Rev. D}
  {\bfseries 42} (1990) 3668}.

\bibitem{fluka}
T.~B{\"o}hlen, F.~Cerutti, M.~Chin, A.~Fass{\`o}, A.~Ferrari, P.~Ortega et~al.,
  \emph{The {FLUKA} code: Developments and challenges for high energy and
  medical applications},
  \href{https://doi.org/10.1016/j.nds.2014.07.049}{\emph{Nuclear Data Sheets}
  {\bfseries 120} (2014) 211 }.

\bibitem{noaa}
\emph{Earth's magnetic field calculator}, {\emph{National Oceanic and
  Atmospheric Administration} (2020)
  \href{https://www.ngdc.noaa.gov/geomag/calculators/magcalc.shtml}{www.ngdc.noaa.gov/geomag/calculators/magcalc.shtml}}.

\bibitem{grieder}
P.~Grieder, \emph{{Cosmic rays at earth: Researcher's reference, manual and
  data book}}. Elsevier, 2001.

\bibitem{g4}
{\scshape GEANT4} collaboration, \emph{{{GEANT4}: A Simulation toolkit}},
  \href{https://doi.org/10.1016/S0168-9002(03)01368-8}{\emph{Nucl. Instrum.
  Meth.} {\bfseries A506} (2003) 250}.

\bibitem{ub_calibration}
{\scshape MicroBooNE} collaboration, \emph{Calibration of the charge and energy
  loss per unit length of the {MicroBooNE} liquid argon time projection chamber
  using muons and protons},
  \href{https://doi.org/10.1088/1748-0221/15/03/p03022}{\emph{JINST} {\bfseries
  15} (2020) P03022} [\href{https://arxiv.org/abs/1907.11736}{{\ttfamily
  1907.11736}}].

\bibitem{mcs}
{\scshape MicroBooNE} collaboration, \emph{Determination of muon momentum in
  the {MicroBooNE} {LArTPC} using an improved model of multiple {Coulomb}
  scattering},
  \href{https://doi.org/10.1088/1748-0221/12/10/p10010}{\emph{JINST} {\bfseries
  12} (2017) P10010} [\href{https://arxiv.org/abs/1703.06187}{{\ttfamily
  1703.06187}}].

\bibitem{deltaicarus}
{\scshape Icarus} collaboration, \emph{{Determination of through-going tracks'
  direction by means of delta-rays in the {ICARUS} liquid argon time projection
  chamber}}, \href{https://doi.org/10.1016/S0168-9002(99)01384-4}{\emph{Nucl.
  Instrum. Meth. A} {\bfseries 449} (2000) 42}.

\bibitem{muonargon}
K.~Ingles, T.~Junk and A.~Marchionni, \emph{Muon energy loss in liquid argon},
  in \emph{Fermilab SIST/GEM Presentations}, 2017,
  \href{https://indico.fnal.gov/event/14933/contributions/28526/}{https://indico.fnal.gov/event/14933/contributions/28526/}.

\bibitem{michel}
{\scshape MicroBooNE} collaboration, \emph{{Michel} electron reconstruction
  using cosmic-ray data from the {MicroBooNE LArTPC}},
  \href{https://doi.org/10.1088/1748-0221/12/09/P09014}{\emph{JINST} {\bfseries
  12} (2017) P09014} [\href{https://arxiv.org/abs/1704.02927}{{\ttfamily
  1704.02927}}].

\bibitem{yifan}
{\scshape MicroBooNE} collaboration, \emph{A method to determine the electric
  field of liquid argon time projection chambers using a {UV} laser system and
  its application in {MicroBooNE}},
  \href{https://doi.org/10.1088/1748-0221/15/07/p07010}{\emph{JINST} {\bfseries
  15} (2020) P07010} [\href{https://arxiv.org/abs/1910.01430}{{\ttfamily
  1910.01430}}].

\bibitem{mooney}
M.~Mooney, \emph{The {MicroBooNE} experiment and the impact of space charge
  effects},  in \emph{{APS DPF proceedings}}, 11, 2015,
  \href{https://arxiv.org/abs/1511.01563}{{\ttfamily 1511.01563}}.

\bibitem{mooney2}
{\scshape MicroBooNE} collaboration, \emph{Measurement of space charge effects
  in the {MicroBooNE} {LArTPC} using cosmic muons},
  \href{https://doi.org/10.1088/1748-0221/15/12/p12037}{\emph{JINST} {\bfseries
  15} (2020) P12037} [\href{https://arxiv.org/abs/2088.09765}{{\ttfamily
  2088.09765}}].

\bibitem{signal1}
{\scshape MicroBooNE} collaboration, \emph{{Ionization electron signal
  processing in single phase {LArTPC}s. Part {I}. Algorithm Description and
  quantitative evaluation with MicroBooNE simulation}},
  \href{https://doi.org/10.1088/1748-0221/13/07/P07006}{\emph{JINST} {\bfseries
  13} (2018) P07006} [\href{https://arxiv.org/abs/1802.08709}{{\ttfamily
  1802.08709}}].

\bibitem{signal2}
{\scshape MicroBooNE} collaboration, \emph{{Ionization electron signal
  processing in single phase {LArTPC}s. Part {II}. Data/simulation comparison
  and performance in MicroBooNE}},
  \href{https://doi.org/10.1088/1748-0221/13/07/P07007}{\emph{JINST} {\bfseries
  13} (2018) P07007} [\href{https://arxiv.org/abs/1804.02583}{{\ttfamily
  1804.02583}}].

\bibitem{pandora}
{\scshape MicroBooNE} collaboration, \emph{The pandora multi-algorithm approach
  to automated pattern recognition of cosmic-ray muon and neutrino events in
  the {MicroBooNE} detector},
  \href{https://doi.org/10.1140/epjc/s10052-017-5481-6}{\emph{Eur. Phys. J. C}
  {\bfseries 78} (2018) 82} [\href{https://arxiv.org/abs/1708.03135}{{\ttfamily
  1708.03135}}].

\bibitem{ub_noise}
{\scshape MicroBooNE} collaboration, \emph{Noise characterization and filtering
  in the {MicroBooNE} liquid argon {TPC}},
  \href{https://doi.org/10.1088/1748-0221/12/08/p08003}{\emph{JINST} {\bfseries
  12} (2017) P08003} [\href{https://arxiv.org/abs/1705.07341}{{\ttfamily
  1705.07341}}].

\bibitem{mucs}
{\scshape MicroBooNE} collaboration, \emph{Measurement of cosmic-ray
  reconstruction efficiencies in the {MicroBooNE} {LArTPC} using a small
  external cosmic-ray counter},
  \href{https://doi.org/10.1088/1748-0221/12/12/p12030}{\emph{JINST} {\bfseries
  12} (2017) P12030} [\href{https://arxiv.org/abs/1707.09903}{{\ttfamily
  1707.09903}}].

\bibitem{previous}
J.~Kremer, M.~Boezio, M.~L. Ambriola, G.~Barbiellini, S.~Bartalucci,
  R.~Bellotti et~al., \emph{Measurements of ground-level muons at two
  geomagnetic locations},
  \href{https://doi.org/10.1103/PhysRevLett.83.4241}{\emph{Phys. Rev. Lett.}
  {\bfseries 83} (1999) 4241}.

\bibitem{muon}
B.~{Mitrica}, R.~{Margineanu}, S.~{Stoica}, M.~{Petcu}, I.~{Brancus}, A.~{Jipa}
  et~al., \emph{A mobile detector for measurements of the atmospheric muon flux
  in underground sites},
  \href{https://doi.org/10.1016/j.nima.2011.07.006}{\emph{Nucl. Intrum. Meth.
  A} {\bfseries 654} (2011) 176}
  [\href{https://arxiv.org/abs/1104.2157}{{\ttfamily 1104.2157}}].

\bibitem{muon2}
D.~Stanca, B.~Mitrica, M.~Petcu, I.~M.~Brancus, A.~Jipa, A.~Haungs et~al.,
  \emph{Measurements of the atmospheric muon flux using a mobile detector based
  on plastic scintillators read-out by optical fibers and {PMTs}},
  \href{https://doi.org/10.1088/1742-6596/409/1/012136}{\emph{J. Phys. Conf.
  Ser} {\bfseries 409} (2013) 012136}.

\bibitem{seasonal}
S.~Cecchini and M.~Spurio, \emph{Atmospheric muons: experimental aspects},
  \href{https://doi.org/10.5194/gi-1-185-2012}{\emph{Geosci. Instrum. Method.
  Data Syst.} {\bfseries 1} (2012) 185}
  [\href{https://arxiv.org/abs/1208.1171}{{\ttfamily 1208.1171}}].

\end{thebibliography}\endgroup

\end{document}